\RequirePackage{fix-cm}
\documentclass{svjour3}                     % onecolumn (standard format)
\smartqed  % flush right qed marks, e.g. at end of proof
\usepackage{graphicx}
\usepackage{amsmath}
\journalname{Journal of Geodesy}
\begin{document}

\title{Gravity Field Mapping Using Laser Coupled Quantum Accelerometers in Space
%\thanks{Grants or other notes
%about the article that should go on the front page should be
%placed here. General acknowledgments should be placed at the end of the article.}
}
%\subtitle{Do you have a subtitle?\\ If so, write it here}

%\titlerunning{Short form of title}        % if too long for running head

\author{T. L\'ev\`eque$^1$         \and
              C. Fallet$^1$         \and
              M. Mandea$^2$         \and
              R. Biancale$^1$\footnote{Deceased, February 2019.}         \and
              J. M. Lemoine$^1$         \and
              S. Tardivel$^1$         \and
              S. Delavault$^1$         \and
              A. Piquereau$^1$         \and
              S. Bourgogne$^3$         \and
              F. Pereira Dos Santos$^4$         \and
              B. Battelier$^5$         \and
              Ph. Bouyer$^5$}

%\authorrunning{Short form of author list} % if too long for running head

\institute{T. L\'ev\`eque \at
               \email{thomas.leveque@cnes.fr}           %  \\
%             \emph{Present address:} of F. Author  %  if needed
           \and
	$^1$ Centre National d'Etudes Spatiales, 18 avenue Edouard Belin, 31400 Toulouse, France.\\
	$^2$ Centre National d'Etudes Spatiales, 2 Place Maurice Quentin, 75001 Paris, France.\\
	$^3$ Stellar Space Studies, 5 Esplanade Compans Caffarelli, 31000 Toulouse, France.\\
	$^4$ LNE-SYRTE, Observatoire de Paris, Universit\'e PSL, CNRS, Sorbonne Universit\'e, 61 Avenue de l'Observatoire, 75014 Paris, France.\\
	$^5$ LP2N, IOGS, CNRS, Universit\'e de Bordeaux, Rue Fran\c{c}ois Mitterrand, 33400 Talence, France.
}

\date{Received: date / Accepted: date}
% The correct dates will be entered by the editor

\maketitle

\begin{abstract}
The emergence of quantum technologies, including cold atom based accelerometers, offers an opportunity to improve the performances of space geodesy missions. In this context, CNES initiated an assessment study called GRICE (GRadiom\'etrie \`a Interf\'erom\`etres quantiques Corr\'el\'es pour l'Espace) in order to evaluate the contribution of cold atom technologies to space geodesy and to the end users of geodetic data. In this paper, we present mission scenario for gravity field mapping based on a long baseline gradiometer. The mission is based on a constellation of two satellites, flying at an altitude of 373 km, each equipped with a cold atom accelerometer with a sensitivity of $6 \times 10^{-10}$~m.s$^{-2}$.$\mathrm{\tau}^{-1/2}$. A laser link measures the distance between the two satellites and couples these two instruments in order to produce a correlated differential acceleration measurement. The main parameters, determining the performances of the payload, have been investigated. We carried out a general study of satellite architecture and simulations of the mission performances in terms of restitution of the gravity field. The simulations show that this concept would give its best performance in terms of monthly gravity fields recovery under 1000~km resolution. In the resolution band between 1000 and 222~km, the improvement of the GRICE gradient approach over the traditional range-rate approach is globally in the order of 10 to 25\%. 

% \subclass{MSC code1 \and MSC code2 \and more}
\end{abstract}

\section{Introduction}
\label{intro}

The knowledge of the gravity field, which reflects the mass distribution of the Earth, enables to investigate the internal structure of our planet and the dynamics of its external fluid layers: atmosphere, oceans, polar caps, hydrosphere. Observations on the gravity field thus contribute to the understanding of the geological and climatic evolutions of our planet~\cite{tapley2019}. Meanwhile, the emergence of quantum technologies, including cold atom based accelerometers, offers an opportunity to improve the performances of space geodesy missions (e.g. GOCE, GRACE, GRACE-FO)~\cite{douch2018}. In this context, we initiated an assessment study~\cite{leveque2018} called GRICE (GRadiom\'etrie \`a Interf\'erom\`etres quantiques Corr\'el\'es pour l'Espace) in order to evaluate the potential contribution of cold atom technologies to space geodesy and to the end users of the geodetic data.

The Earth is a complex system composed of an internal structure (core, mantle, crust) and its external fluid layers (atmosphere, oceans, polar caps, hydrosphere). This system is submitted to permanent interactions between its components over different scales of time and space. These mass transfers result in temporal variations of the gravity field, followed since 2002 with a global coverage by the GRACE mission~\cite{tapley2004}. On the other hand, the undulations of the so-called "static" field, mapped by the GOCE mission~\cite{drinkwater2007}, up to about 85 km of resolution~\cite{bruinsma2014} provide information on the internal distribution of the masses of the planet and its evolution at the geological scales. It also provides the equipotential reference surface, called geoid, to which the altitudes and the dynamic topography of the oceans relate. The use of the space gravimetry data is a real challenge because of the wealth of its signal. Indeed, its exploitation requires to identify the contributions of superimposed sources in the measurement of the field. The GOCE mission has shown the relevance of the gravity gradient measurement to identify the different sources based on their geometric characteristics. More generally, the method used for the separation of superimposed contributions to the gravity field is based on the identification of specific spatial and temporal patterns. These are combined to observations and models on the individual components of the system (satellite data of soil moisture, altimetry, soil deformations, ...). It has been shown that these analyzes benefit from an increase of the spatial resolution, particularly for discriminating the deformations of the solid Earth from that of the water cycle in the fluid envelopes. It can also contribute to distinguish the signal of the postglacial rebound from that of the ice melt in the polar areas. An increase in spatial resolution also enables to reduce the "spatial leakage" at the boundaries between areas of different variability, and therefore, to better estimate their mass balances. Thus, the scientific needs for the future space missions, in terms of measurement of the Earth's gravity field, are first, the continuation of long-term observations, which are crucial for the study of the climate change. Second, it is necessary to increase the spatial resolution of this kind of mission in order to close mass balances at all scales of time and space~\cite{pail2015}. Finally, the scientific community would benefit from a mission determining the temporal variations of the gravity field with increased performances compared to GRACE.

\begin{figure*}
\centering  \resizebox{12cm}{!}{\includegraphics[angle=-90,origin=c]{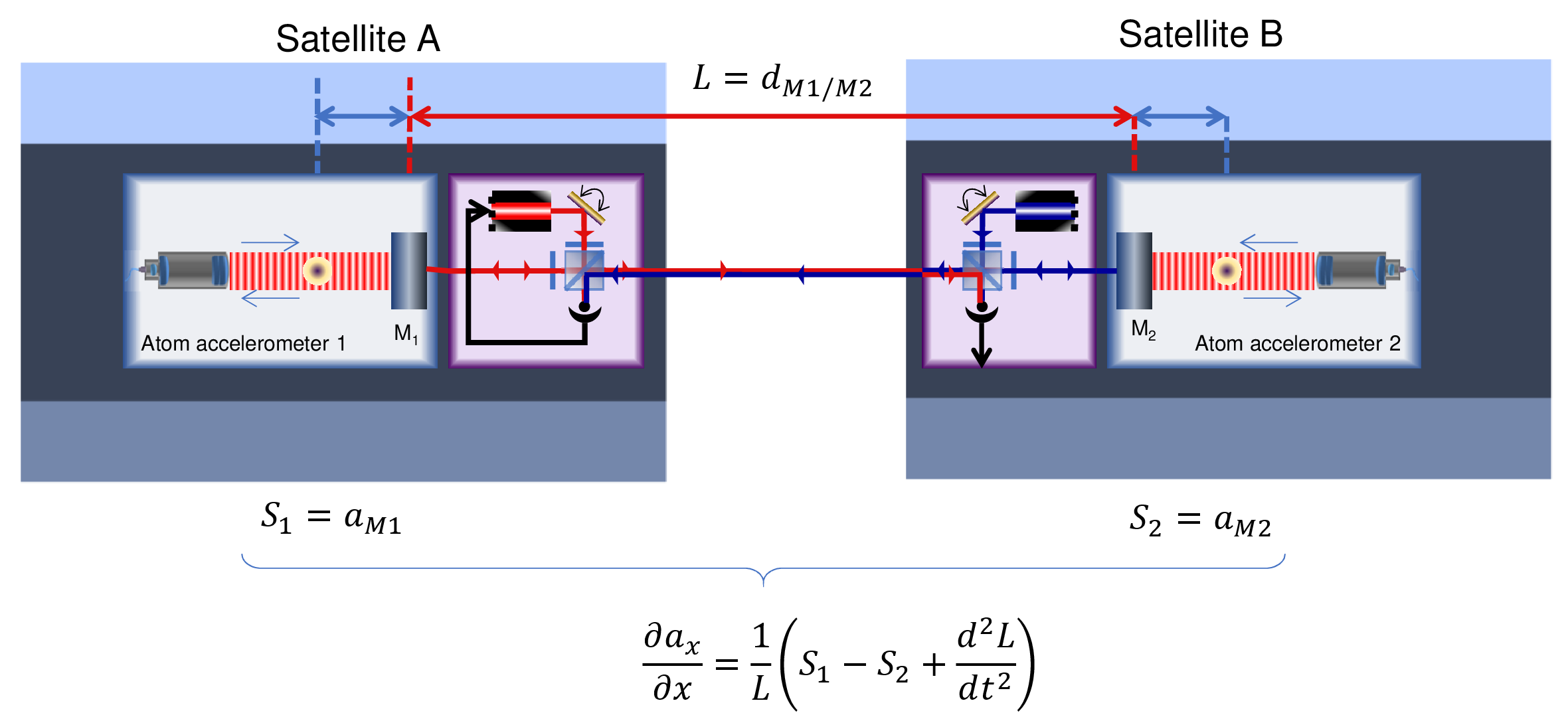}}
\caption{Mission scenario involving twin satellites each equipped with a cold atom accelerometer. A laser telemeter is used to monitor the distance between the two satellites and couples the measurements of the two atom interferometers.}
\label{setup}
\end{figure*}

Here, we present a space mission concept for mapping the Earth's gravity field based on a long baseline gravity gradiometer. The mission uses a constellation of two satellites each equipped with a cold atom accelerometer. A laser link enables to measure the distance between the two satellites and then couples these two instruments in order to produce a differential acceleration measurement. This mission, composed of one pair of satellites, aims to improve the space resolution of Earth’s gravity field. A complete analysis of this concept was carried out in order to determine the performances of the instrument and the main mission parameters. A first satellite layout definition has been realized and the performances of the mission in terms of restitution of the Earth's gravity field have been evaluated. A 5-year mission lifetime is envisioned.

The article is structured as follows. In the first part, we present the instrument concept for a future gravity mission which enables to realize a long baseline gravity gradient measurement on a twin satellite constellation. The main performances of the payload are presented. The second part is dedicated to the definition of the mission parameters. We then determine the orbit of the satellites in order to maximize the sensitivity to the gravity field and to obtain an optimal coverage of the Earth. The needs in terms of on board energy (solar panel, battery) and propulsion are deducted from the orbit. This analysis enables to end up with a preliminary design of the satellite platform. The last section is dedicated to a numerical simulation in order to retrieve the mission performances that could be achieved in terms of restitution of the Earth's gravity field.

\section{Satellite Payload}
\label{sec:payload}

One of the main scientific needs, in terms of gravity field mapping from space, are related to an increase of the spatial resolution of the measurements. Moreover, stability and accuracy would ease long time data treatments and combinations with other data types. Finally, the gravity gradient is the most relevant observable to increase the spatial resolution of the mission as it will be discussed in section~\ref{comparison}. In this context, the development of cold atom based accelerometers offers an opportunity to improve the performances of space geodesy missions. Indeed, these instruments provide stable and accurate measurements and are particularly adapted to differential mode operations needed for the gravity gradient measurement~\cite{snadden1998}. In this part, we describe the use of quantum technologies in a specific configuration that would enable an increase of the performances of space geodesy missions.

\subsection{General description}

Two main space mission concepts have been investigated for mapping the Earth's gravity field using cold atom instruments~\cite{carraz2014,trimeche2019,chiow2015,hogan2016,abrykosov2019,migliaccio2019}. The first one, which has been proposed in~\cite{carraz2014} and studied in~\cite{trimeche2019}, is based on a 3-axis atomic gradiometer of 0.5-meter baseline on board a single satellite. It consists in a GOCE-like mission at a low altitude of 239 km. This instrument would enable to achieve a sensitivity of 5~mE.Hz$^{-1/2}$, and then a two-fold improvement on the gravity field recovery for degrees above 50~\cite{trimeche2019}. The second concept, which has been proposed in~\cite{chiow2015}, involves a long baseline gradiometric measurement between two satellites~\cite{keller2005}. A similar twin-satellite concept has also been proposed for gravitational wave detection from space in~\cite{hogan2016}.

In this paper, we investigate a twin-satellite mission concept inspired by~\cite{chiow2015}. The instrument, depicted on the Figure~\ref{setup}, is based on a composite gradiometric measurement. The mission uses a constellation of two satellites each equipped with a cold atom accelerometer placed at its center of mass. A laser link monitor the relative displacement between the two satellites in order to perform a differential measurement between the two distant accelerometers and reject the common-mode noises. The absence of drifts on the atomic measurements allows for the accurate determination of the differential inertial acceleration between the two satellites out of the combination of these accelerometric data. The mission payload is then functionally close to the GRACE and GRACE follow-on missions~\cite{kornfeld2019}. This measurement concept offers the prospect of improved mission performances as it will be shown in section~\ref{sec:simulation}.

In this instrument concept, each atom accelerometer measures the acceleration, $a_{M1}$ and $a_{M2}$, between an inertial frame, defined by the free falling atomic cloud, and the frame of its satellite, materialized by the retroreflection mirror of the Raman beam, which is rigidly attached to the satellite. We define the signals delivered by the atom accelerometers $S_{1,2}$ as:

\begin{equation}
S_1=a_{M1} ; S_2=a_{M2}.
\end{equation}

The laser link provides a simultaneous telemetry measurement between the two frames of the satellites, i.e. between the two mirrors of the atom accelerometers:

\begin{equation}
L=d_{M1/M2}.
\end{equation}

Thus, the combination of the two acceleration measurements with the second derivative of the telemetry signal enables to determine the gradient of acceleration between the two distant atomic clouds along the track of the satellites ($a_x$):

\begin{equation}
\label{gradient}
\frac{\partial a_x}{\partial x}=\frac{1}{L} \left(S_1-S_2+\frac{d^2L}{dt^2} \right).
\end{equation}

In order to provide the best performances for the complete instrument, the derivative treatment of the telemetry signal has to be included in the data processing as it is crucial to guarantee the coupling between the three instruments.

The gravity gradient measurement expressed in Equation (\ref{gradient}) becomes more sensitive when the baseline of the instrument (i.e. the distance between the two accelerometers) increases. This payload configuration, which distributes the atom accelerometers on two satellites, allows a significant increase of the baseline compared to an instrument that would have to stand on a unique platform. In this constellation, each satellite is implemented at low orbit in a nadir pointing mode and then submitted to an orbital rotation $\Omega_{orb}$ of $1.16 \times 10^{-3}$~rad.s$^{-1}$ at an altitude of 373~km.

\begin{table*}
% table caption is above the table
\caption{Main parameters of the cold atom accelerometer envisioned for the GRICE mission.}
\label{tab1}       % Give a unique label
% For LaTeX tables use
\centering
\begin{tabular}{llll}
\hline\noalign{\smallskip}
Parameter & Notation & Value & Unit  \\
\noalign{\smallskip}\hline\noalign{\smallskip}
Interaction time&$T$&0.5&s \\
Contrast&$C$&0.43&- \\
Preparation time &  $t_{prep}$ &1&s \\
Cycle duration& $t_c$&2&s \\
Detected atom number& $N_{det}$ & $5 \times 10^5$ &- \\
Atomic temperature& $\theta$ & $5 \times 10^{-11}$ & K \\
Initial cloud size& $\sigma_0$ & $1 \times 10^{-4}$ & m \\
Accelerometer sensitivity& $\sigma_a$ & $6 \times 10^{-10}$ & m.s$^{-2}$.$\tau^{-1/2}$ \\
\noalign{\smallskip}\hline
\end{tabular}
\end{table*}

\subsection{Atom accelerometers}

\begin{figure}
\centering \resizebox{8cm}{!}{\includegraphics[angle=-90,origin=c]{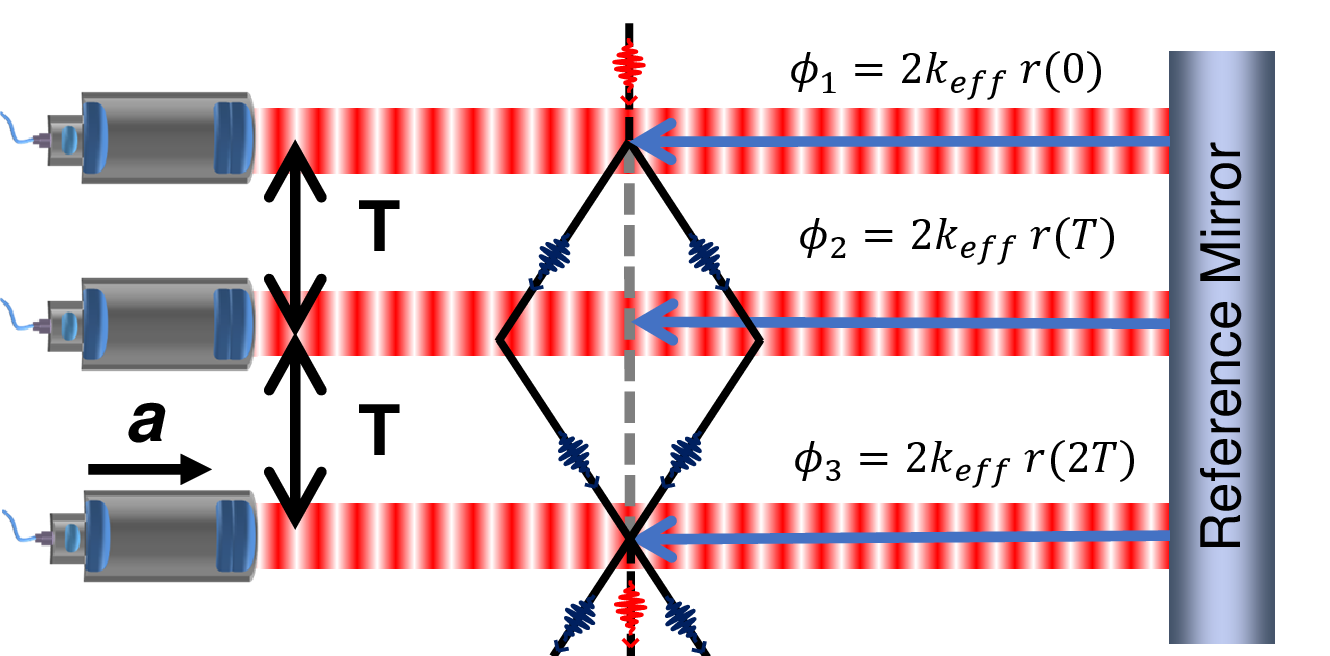}}
%[angle=-90,origin=c] 
%\includegraphics[width=0.75\textwidth]{interferometer.eps}
\caption{Scheme of a 3 pulses atom interferometer using Raman lasers in a retro-reflected configuration. The total interaction time is 2T. The interferometer is sensitive to the acceleration of the cold atoms with respect to the retro-reflection mirror in the direction of the Raman lasers.}
\label{interferometer}
\end{figure}

For this mission, we consider atom accelerometers based on sources of laser cooled Rubidium atoms manipulated by Raman transitions (Figure~\ref{interferometer}). Rubidium atoms are first prepared from a vapor to form an ultra-cold sample, using atom chip techniques~\cite{schuldt2015,becker2018,muntinga2013}. The atomic cloud is then manipulated, so as to move it away from the surface of the chip~\cite{amri2018} and collimate it via magnetic lensing~\cite{kovachy2015}. Then, the free falling Rubidium cloud interacts successively three times with a unique pair of retro-reflected Raman beams, which acts on matter waves as beam splitters or mirrors. This creates an interferometer of $2T$ total interaction time in a so called double-diffraction configuration~\cite{leveque2009}. This interferometer scheme is particularly adapted to space instruments. Indeed, for inertial sensors in zero-gravity environments, the Doppler effect cannot be used to select one or the other effective Raman transitions in the current retro-reflected configuration needed for accuracy. The atomic phase shift is then obtained from the population in each output port of the interferometer, which is measured by a fluorescence technique. The atomic phase shift $\Delta \Phi$ is given by:

\begin{equation}
\Delta \Phi=\Phi_1-2\Phi_2+\Phi_3,
\end{equation}

where $\Phi_{1,2,3}$ is the local Raman laser phase seen by the atom along their classical path. This phase shift is then related to the acceleration between the atomic cloud and the equiphases of the Raman laser, along its propagation direction, by:

\begin{equation}
\Delta \Phi=2k_{eff}  a T^2,
\end{equation}

Where $k_{eff}$ corresponds to the effective wave vector of the Raman beam. The sensitivity of the acceleration measurement is related to the interaction time ($T$), the fringe contrast ($C$), the duration of the measurement cycle ($t_c$) and the detection noise. In our study, we consider that the determination of the output phase shift of the interferometer is limited by the quantum projection noise~\cite{gauguet2009}. Indeed, other noise contributions can be reduced down to the quantum projection noise limit when using atom interferometers in differential mode~\cite{sorrentino2014}. Moreover, the double-diffraction configuration further suppresses the influence of many systematics and noise sources with respect to single-k diffraction interferometers. The quantum projection noise is linked to the atom number ($N_{det}$) at the output of the interferometer, so that the acceleration sensitivity ($\sigma_a$) is given by:

\begin{equation}
\sigma_a=\frac{1}{k_{eff}T^2} \frac{1}{C \sqrt{N_{det}}} \frac{1}{\sqrt{t_c}}.
\end{equation}

The fringe contrast depends on the atomic cloud temperature and decreases with the rotation of the satellite and the interaction time as described in~\cite{barrett2016}. Indeed, in the scenario that we consider, the sensitivity axis of the instruments is orientated along the line of sight of the laser ranging instrument while the spacecraft operates in a nadir pointing mode. The rotation of the instrument during the orbit causes a loss of contrast due to the separation of the wave-packet trajectories and the resulting imperfect overlap during the final light pulse. The contrast is then reduced due to the rotation of the spacecraft in the orbital plane. In the case of an atom interferometer in double-diffraction configuration, this loss of contrast is given by:

\begin{equation}
C \propto \exp\left[-\left( k_{eff} \sigma_r (2T) T\right)^2 (\Omega_x^2+\Omega_y^2 )\right]
\end{equation}

where $\sigma_r(t)$ is the size of the atomic wavepacket given by:

\begin{equation}
\sigma_r (t)=\sqrt{\left(\frac{\hbar}{\sigma_p}\right)^2 + \left(\frac{\sigma_p t}{M}\right)^2}  \textrm{and } \sigma_p=\sqrt{2k_B M\theta}
\end{equation}

In this expression, $k_B$ corresponds to the Boltzmann constant, M to the mass of the $^{87}$Rb atoms and $\theta$ to the temperature of the atomic cloud. In order to evaluate the atom accelerometer performances needed for this mission, we list the key parameters of the instrument. In the Table~\ref{tab1}, we report a set of parameters, achievable by existing technologies, which enables to reach a sensitivity on the acceleration measurement of $6 \times 10^{-10}$~m.s$^{-2}$.$\tau^{-1/2}$, where $\tau$ is the averaging time. This sensitivity corresponds to a noise of 2.3~mrad per shot on the phase measurement of the atom interferometer. The interaction time T of 0.5~s is chosen so as to ensure a high sensitivity to accelerations while limiting the loss of contrast induced by nadir rotation. This set of parameters guarantees the existence of a signal at the output of the atom interferometer independently of any rotation compensation system.

\subsection{Laser ranging instrument}

The laser ranging instrument is used to determine the relative displacement of the two atom accelerometers in order to retrieve the gravity gradient measurement along the track of the satellites. As the measurement is realized between the two retroreflection Raman mirrors of the accelerometers, the gravity gradient measurement extracted from the combination of atomic accelerometer and LRI data is, in principle, insensitive to non-gravitational forces acting on the satellites. The laser ranging instrument considered in our study, is based on a heterodyne optical interferometer working at a wavelength $\lambda$ of 1.5~$\mu$m. The principle of this instrument (Figure~\ref{setup}) is similar to the one used in the GRACE FO mission~\cite{sheard2012,sanjuan2015}.
A first laser L1, on board the satellite A, is frequency locked on a spectroscopic reference. This laser is sent to the satellite B after a retroreflection on the local Raman mirror. On board the satellite B, the laser L1 is received and superimposed to a local laser L2 in order to form a beat note on a fast photodiode at a few tenth of MHz. This signal is used to lock the phase of the laser L2 onto the laser L1. The laser L2 is then sent back to the satellite A where its phase is measured with respect to the local arm of the laser L1 through a beat note technique. The intersatellite displacement $\delta_{L}$ is derived from the phase shift $\delta \Phi$ of the beat note:

\begin{equation}
\delta_{L}=\frac{\lambda}{2 \pi} \delta \Phi
\end{equation}

For generating the composite gradient of acceleration, the signals coming from the laser ranging and from the accelerometers have to be recorded simultaneously. The ranging signal is then differentiated twice and filtered taking into account the sensitivity function of the atom accelerometers~\cite{cheinet2008,lautier2014}. This treatment is an important aspect which has to be included in the data processing in order to guarantee an optimal coupling between the instruments. This laser ranging system also integrates a steering mirror in order to perform the fine pointing of the beam on the distant satellite~\cite{koch2018}.

\begin{table}
% table caption is above the table
\caption{Main requirements for the intersatellite laser ranging instrument.}
\label{tab2}       % Give a unique label
% For LaTeX tables use
\begin{tabular}{lll}
\hline\noalign{\smallskip}
Parameter&Value&Unit\\
\noalign{\smallskip}\hline\noalign{\smallskip}
Sensitivity&40 & nm. Hz$^{-1/2}$ \\
Intersatellite distance& $\approx$ 100 &km \\
Displacement range& $\pm$1 & km \\
Relative velocity range& $\pm$0.5 & m.s$^{-1}$ \\
Relative acceleration range & $\pm 8 \times 10^{-4}$ & m.s$^{-2}$ \\
\noalign{\smallskip}\hline
\end{tabular}
\end{table}

In the frame of the proposed mission, the laser ranging instrument has to be able to operate following the relative motion of the satellites along track which corresponds to the ranges reported in the Table~\ref{tab2}. The performance required for the laser ranging measurement is of 40~nm.Hz$^{-1/2}$ in order not to limit the performance of the atom accelerometer as it is shown in the section~\ref{sec:performances}. This performance is close to the in orbit performance of the GRACE-FO mission laser ranging instrument~\cite{abich2019}. Nevertheless, low frequency increase of the laser ranging instrument noise below 40~mHz is not considered here as the overall noise of the gravity gradient measurement will be dominated by the atom accelerometers contribution in this frequency domain.

In order to guarantee this level of performances, the relative frequency stability of the reference laser has to be better than 10$^{-13}$~Hz$^{-1/2}$. This level of performances is currently achievable at $\lambda$=1.5~$\mu$m by frequency locking a laser either on a high finesse Fabry-Perot cavity~\cite{argence2012} or on saturated absorption spectroscopy in Iodine~\cite{philippe2017}.

\subsection{Instrument performances}
\label{sec:performances}

\begin{figure}
\centering \resizebox{8cm}{!}{\includegraphics{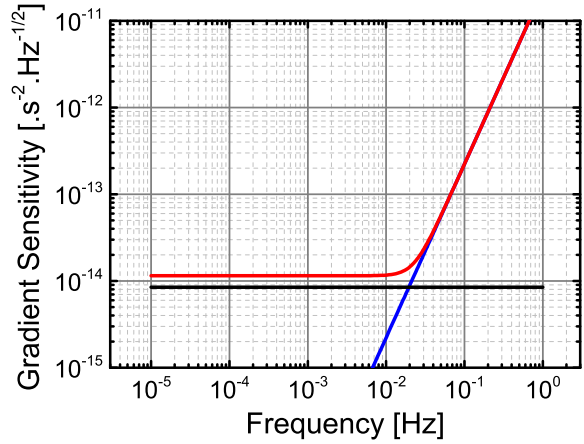}} 
\caption{Amplitude spectral density of noise of the complete instrument (red) expressed in sensitivity to the gravity gradient along the track. Individual noise contributions of atom accelerometers (black) and laser ranging (blue) to the complete instrument noise are also represented.}
\label{instrument_performances}
\end{figure}

The gravity gradient measurement is retrieved from the combination of the acceleration data given by the atomic instruments and of the laser ranging instrument. The laser ranging instrument, which provides a distance measurement, is derived twice in order to determine the relative acceleration between the two satellites. This derivation results in an increase of the measurement noise with the frequency. Finally, the sensitivity of the combined instrument is limited by the atom accelerometers from 10$^{-5}$ to 10$^{-2}$~Hz and by the laser ranging instrument noise from 10$^{-2}$ to 1~Hz. The sensitivity of the instrument to the gravity field gradient is reported on the Figure~\ref{instrument_performances}.

\subsection{Impact of the satellite motion on the measurement}

In this part, we determine how to guarantee the performance of the constellation when the satellites are in orbit. To do so, we quantify the impact of the satellite pointing errors on the instrument measurement. We also quantify the impact of the nadir rotation on the phase shift of the atom interferometer.

\subsubsection{Non gravitational accelerations}

Both satellites are submitted to non-gravitational forces affecting the three directions. These forces are mainly due to drag effect of the residual atmosphere in low orbit. In this part, we evaluate the impact of non-gravitational accelerations on the gravity gradient measurement.

\begin{figure}
\centering \resizebox{8cm}{!}{\includegraphics{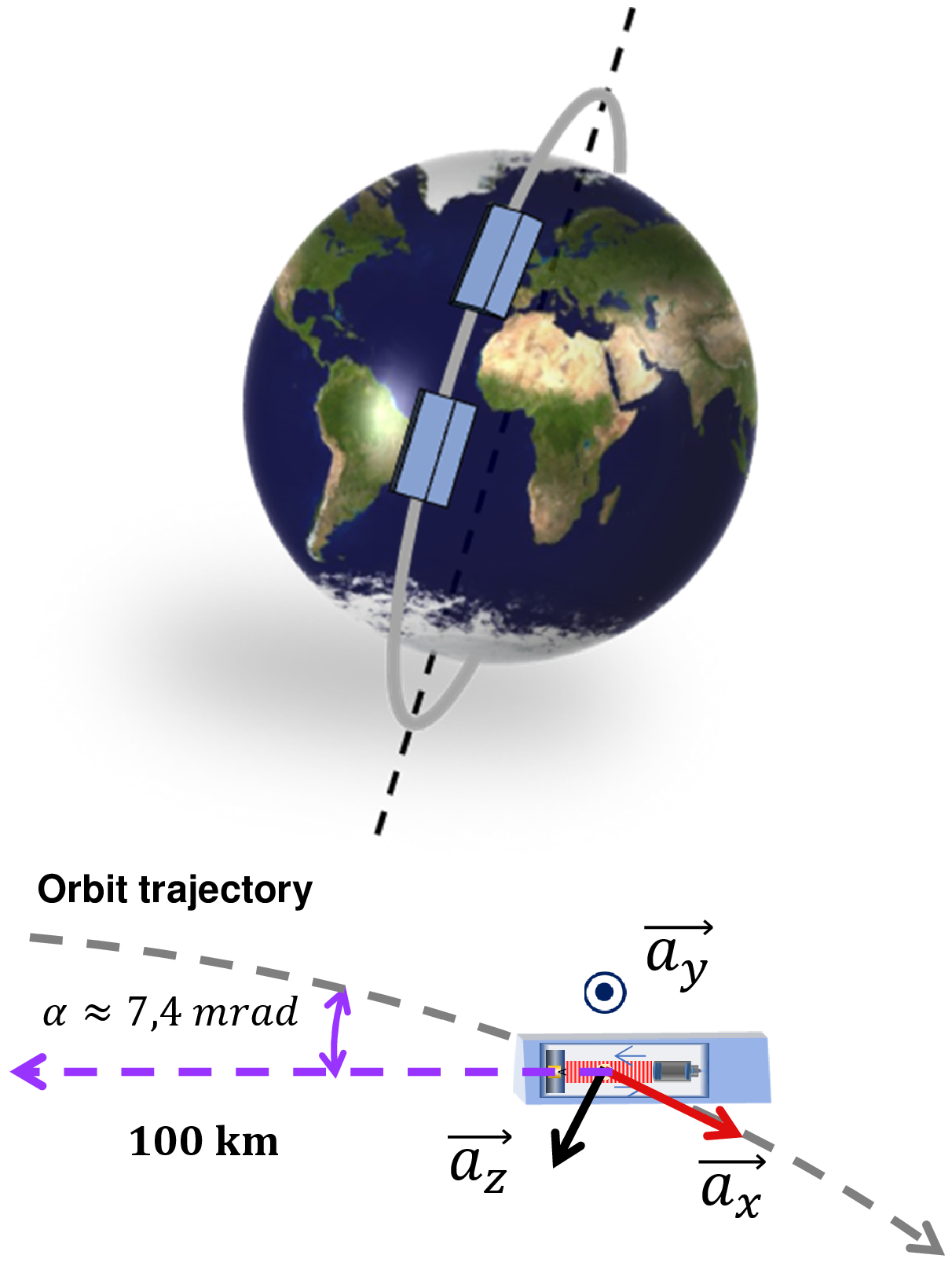}} 
\caption{Description of the main axis of the satellite and laser pointing direction with respect to the satellite trajectory when the two satellites are perfectly aligned.}
\label{mission}
\end{figure}

In order to determine the order of magnitude of the non-gravitational accelerations affecting the satellites, we consider the acceleration levels recorded by the GOCE mission. These levels are not directly comparable to our mission scenario as the altitude of GOCE is lower (260~km). Nevertheless, they can be considered as a worse case for our study as GOCE was submitted to a stronger interaction with the residual atmosphere. The typical levels used for our calculations are reported in Table~\ref{tab3}.

\begin{table}[!h]
% table caption is above the table
\caption{Mean values and standard deviation of the non-gravitational accelerations affecting the GOCE satellite along each direction, for the month of April 2012.}
\label{tab3}       % Give a unique label
% For LaTeX tables use
\centering
\begin{tabular}{lll}
\hline\noalign{\smallskip}
Component&Acceleration&Standard deviation \\
 & mean value [m.s$^{-2}$]&[m.s$^{-2}$]\\
\noalign{\smallskip}\hline\noalign{\smallskip}
$a_x$ (along track)&	$-4.45 \times 10^{-6}$ & $1.4 \times 10^{-6}$\\
$a_y$ (cross track)& $1.16 \times 10^{-7}$ & $1.5 \times 10^{-7}$\\
$a_z$ (Nadir)& $-2.14 \times 10^{-8}$ & $1.7 \times 10^{-8}$\\
\noalign{\smallskip}\hline
\end{tabular}
\end{table}

In the following analysis, we consider that the satellite attitude is controlled within $\pm$4~mrad while the absolute attitude is measured with a sensitivity of $\pm$0.1~mrad. The performances considered here are those of the GRACE mission~\cite{herman2004} and are therefore technically very conservative.

When the two satellites are perfectly aligned as shown in Figure~\ref{mission}, the measurement axis of the laser link is coincident with the atom accelerometers axis. In this case, the non-gravitational accelerations affecting the sensitivity axis are, in principle, perfectly cancelled due to common mode rejection. However, when one of the satellites is affected by a pointing error, the measuring axis of the atom accelerometer is affected by a projection of transverse non-gravitational forces that are not measured by the laser link, and thus induce noise on the acceleration gradient measurement. In the Table~\ref{tab4}, we report the mean values of the parasitic accelerations induced on the atom accelerometer measurement considering satellite pointing errors of 4~mrad along the three axis of rotation.

\begin{table}[!h]
% table caption is above the table
\caption{Mean values of the parasitic accelerations induced on the atom accelerometers by a satellite pointing error of 4~mrad.}
\label{tab4}       % Give a unique label
% For LaTeX tables use
\centering
\begin{tabular}{lll}
\hline\noalign{\smallskip}
Attitude motion&Parasitic acceleration&Standard deviation\\
 &mean value [m.s$^{-2}$]&[m.s$^{-2}$]\\
\noalign{\smallskip}\hline\noalign{\smallskip}
Roll&No effect&-\\
Pitch& $2.5 \times 10^{-10}$ & $6.7 \times 10^{-11}$\\
Yaw& $5 \times 10^{-10}$ & $6.1 \times 10^{-10}$\\
\noalign{\smallskip}\hline
\end{tabular}
\end{table}

In all cases, the parasitic accelerations induced by the non-gravitational accelerations on the atomic interferometer are lower than the sensitivity level of the instrument ($6 \times 10^{-10}$~m.s$^{-2}$). These therefore have no impact on the measurements performed by the constellation. The conservative specifications considered for the satellite attitude control are then sufficient.

\begin{table*}
% table caption is above the table
\caption{Dominant terms induced on the atom interferometer phase shift by the rotation of the satellite. The numerical application is given for $T_{xx}$ = -1.299~s$^{-2}$ which corresponds to the gravity gradient component at an altitude of 373~km.}
\label{tab5}       % Give a unique label
% For LaTeX tables use
\centering
\begin{tabular}{lll}
\hline\noalign{\smallskip}
Source parameter&Phase shift&Numerical application\\
\noalign{\smallskip}\hline\noalign{\smallskip}
$v_x$&	$2k_{eff} v_x (T_{xx}-3(\Omega_y^2+\Omega_z^2))T^3+ O[T]^5$	&21 rad/(m.s$^{-1}$)\\
$v_y$&	$4k_{eff} v_y \Omega_z T^2+O[T]^3$	&0\\
$v_z$&	$-4k_{eff} v_z \Omega_yT^2+O[T]^3$	& $1.8 \times 10^4$ rad/(m.s$^{-1}$)\\
$x_A$&	$2k_{eff} x_A (T_{xx}+\Omega_y^2+\Omega_z^2)T^2+O[T]^3$ & 0.4 rad.m$^{-1}$\\
$y_A$&	$-2k_{eff} y_A \Omega_x \Omega_y  T^2+O[T]^3$	&0\\
$z_A$&	$-2k_{eff} z_A \Omega_x \Omega_z T^2+O[T]^3$	&0\\
\noalign{\smallskip}\hline
\end{tabular}
\end{table*}

\subsubsection{Nadir rotation}

Rotations induce, in addition to a loss of contrast, two types of parasitic contributions on the phase shift of the atomic interferometer~\cite{trimeche2019}. The first is the Coriolis acceleration (Sagnac effect), which scales linearly with the angular velocity ($\Omega$) and the velocity of the atoms in the satellite frame, at the first order. The second is the centrifugal acceleration, which scales like $\Omega^2$, and which depends on the decentering between the atomic cloud and the center of gravity of the satellite at the time of the interferometer. In addition, higher order contributions couple these two inertial forces, in particular with gravity gradients. The impact of these two parasitic contributions is thus related to the control of the parameters of the atomic source (position and velocity) and their fluctuations~\cite{rudolph2016}. This phase shift is added to the acceleration phase shift and thus induces a bias on the measurement of the instrument. In Table~\ref{tab5}, we report the analytical expression of the dominant terms induced on the atom interferometer phase shift by the rotation of the satellite as a function of each parameter of the atomic source. We perform a numerical application of the expression in order to quantify the impact of the nadir rotation on the phase shift of the atom interferometer, and the sensitivity to the atomic velocity and position fluctuations. In this numerical application, we consider the ideal case where $\Omega_y = \Omega_{orb}$, $\Omega_x = 0$ and $\Omega_z = 0$.

The main contribution of the nadir rotation on the atom interferometer phase shift comes from the Sagnac effect induced by the velocity in the $z$ direction. As the sensitivity of the atom accelerometer is at the level of 2.3~mrad per shot, a technical solution for limiting the impact of the rotations on the measurement would be necessary. Several solutions can be envisaged. First, an active control of the center of mass of the satellite would be performed in order to cancel the effect of long term variations of the relative decentering between the center of mass of the spacecraft and the atomic cloud. Second, a high performance fiber gyroscope would be implemented on board in order to measure the rotation and cancel its contribution on the output phase shift, assuming a simultaneous measurement of the residual atomic cloud velocity. Finally, the Sagnac phase and centrifugal acceleration could be limited by an active rotation of the reference mirror in order to compensate for orbital rotation~\cite{lan2012}. This last solution would in addition restore the full contrast of the interferometer and improve its sensitivity.

\begin{figure}
\centering \resizebox{8.5cm}{!}{\includegraphics[angle=-90,origin=c]{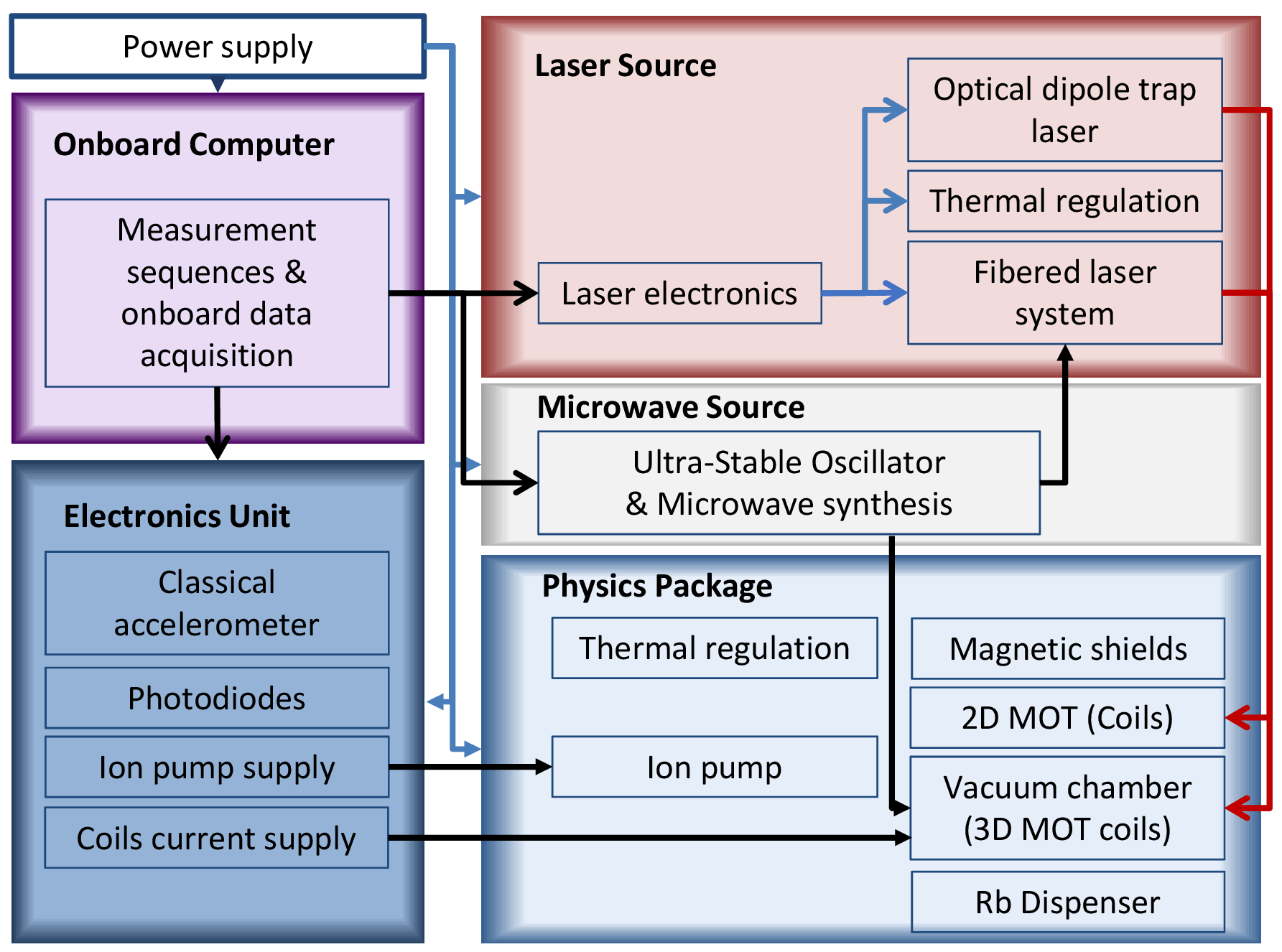}} 
%\centering \includegraphics[width=0.75\textwidth]{diagram.eps}
\caption{Functional diagram of the atom accelerometer.}
\label{diagram}
\end{figure}

\subsection{Instrument architecture}

The atom accelerometer is divided into several subsystems that can be analyzed separately to establish a payload layout plan. The function of the different subsystems depicted on Figure~\ref{diagram} is described below:

\paragraph{The Physics Package (PP)} consists of a titanium vacuum chamber in which the Rubidium atoms are manipulated. It also includes the pumping system to keep the chamber under ultrahigh vacuum, the Rubidium dispenser, and the coils which enable to generate the magnetic fields during the instrument operation. The physics package is surrounded by two layers of magnetic shields.

\paragraph{The Laser Source (LS)} has three main functions during the accelerometer operation: the cold atom preparation, the Raman pulses generation, and the atomic state detection. It is made of a main laser subsystem for generating 2D MOT, 3D MOT, selection, detection and Raman beams. It also contains a more powerful auxiliary laser system dedicated to the realization of an optical dipole trap which would be necessary for manipulating and preparing the cold atom sample before the interferometer sequence.

\paragraph{The Microwave Source (MS)} generates the 6,8~GHz reference signal necessary for the phase/frequency locking of the lasers. This subsystem also enables to generate the electromagnetic signals for driving microwave transitions during the operation of the instrument. It integrates an Ultra-Stable Oscillator (USO) and a microwave synthesis system.

\paragraph{The Electronics Unit (EU)} contains all the electronic subsystems to ensure the operation of the PP. In particular, it includes the ion pump power supply, the magnetic field coil control system, the photodiode acquisition electronics and an accelerometer for hybrid measurements.

\paragraph{The On board Computer (OC)} allows to control the instrument by driving the measurement sequences (atomic cooling, Raman pulses and output phase shift measurement). It also enables to perform the on board data acquisition and treatment.

In the Table~\ref{tab6}, we give an estimation of the power, mass, and volume budget of the payload on board each satellite. This estimation is based on the actual values of the PHARAO mission~\cite{cacciapuoti2009,leveque2015} including dedicated upgrades of technologies~\cite{leveque2014}. The overall budget for the complete payload, including the atom accelerometer and the laser transponder for intersatellite displacement measurement, amounts to a power consumption of 217~W and a mass of 108~kg contained in a volume of 144~l.

\begin{table}
% table caption is above the table
\caption{Mass and power budget for the payload on board each satellite.}
\label{tab6}       % Give a unique label
% For LaTeX tables use
\begin{tabular}{llll}
\hline\noalign{\smallskip}
 &Mass [kg]&	Power [W]&	Volume [l]\\
\noalign{\smallskip}\hline\noalign{\smallskip}
Physics Package&	45&	8&	82\\
Electronics&	5&	57&	5\\
Laser source&	25&	96&	33\\
Microwave source&	7&	26&	9\\
Onboard computer&	6&	10&	7\\
Laser link&	20&	20&	8\\
% & & & \\
\noalign{\smallskip}\hline\noalign{\smallskip}
Total&	108&	217&	144\\
\noalign{\smallskip}\hline
\end{tabular}
\end{table}

\section{Space mission concept}
\label{sec:mission}

The orbit of satellites is determined in order to maximize the sensitivity to the gravity field, via the altitude, while performing an optimal coverage of the Earth. From the orbit, we derive the needs in terms of on board energy (solar panel, battery) and propulsion for keeping the orbit during the mission. This analysis enables to end up with a preliminary design of the satellite platform.

\subsection{Mission analysis}

\subsubsection{Choosing the orbit}

In a low–low satellite-to-satellite tracking mission dedicated to gravity recovery, the baseline of the instrument (i.e. the separation between the satellites) contributes to determine the shortest measurable wavelength of the gravity field. Thus, in our study, an intersatellite separation of 100~km has been chosen in order to increase the sensitivity to the short-wavelength part. Moreover, so as to optimize the gravity field restitution, the orbit of the two satellites has to fulfill the following requirements:

\paragraph{Inclination:} a full coverage of the ice polar caps is necessary as their gravity observations are of major scientific interest.

\paragraph{Altitude:} the altitude should be as low as possible to maximize the sensitivity of the constellation to the high spatial frequencies of the gravity field while limiting the effect of the residual atmosphere. The altitude must be chosen in the range 320-420~km. The altitude must also be constant. An altitude of 373~km was chosen, based on initial analytical simulations of performance for the recovery of the gravity field. 

\paragraph{Groundtrack:} the orbit is optimized to have an inter-track lower than 10~km and a regular ground track on a weekly and monthly basis. This requires a very long phasing cycle of about a year, with two sub-cycles: one of about 30 days and the other of about a week. This specification enables to recover a homogeneous groundtrack densification along the mission lifetime. This homogeneous distribution of the groundtracks allows an optimal gravity map restitution on several time scales.

These constraints set several orbital parameters: an inclination of  90$^\circ$ since the orbit is polar, a frozen eccentricity since the orbit shall be quasi–circular (constant altitude). Hence the argument of the perigee is set to 90$^\circ$ and the eccentricity value will be directly determined by the specific altitude chosen. The altitude and the right ascension of the ascending node (RAAN) are determined taking into account the constraints of the mission. The final orbit parameters are indicated in Table~\ref{orbit_simu}.

\begin{table}[!h]
% table caption is above the table
\caption{Orbit parameters of the mission.}
\label{orbit_simu}       % Give a unique label
% For LaTeX tables use
\centering
\begin{tabular}{lll}
\hline\noalign{\smallskip}
Parameter &	Value &	 Unit \\
\noalign{\smallskip}\hline\noalign{\smallskip}
Semi-major axis & 6751 440.4 & m\\
Eccentricity & 0.0011 & -\\
Inclination & 90 & deg.\\
RAAN & 0 & deg.\\
Argument of perigee & 90 & deg.\\
% & & & \\
\noalign{\smallskip}\hline
\end{tabular}
\end{table}

\subsubsection{$\Delta V$ budget and maneuvers}

During the mission, the two satellites will be controlled in altitude and in phasing. The orbital plane will not be controlled as the drifts will remain small. The main change in velocity ($\Delta V$) cost then comes from the effects of atmospheric drag. In this mission, the worst case of area to mass ratio for the satellites is estimated at $2.9 \times 10^{-2}$~m$^2$.kg$^{-1}$ according to the GRACE satellite parameters~\cite{herman2004}. From this parameter,  we compute rates of semi-major axis decay between 50~m/day to 300~m/day depending on solar activity. The worst case scenario yields a required $\Delta V$ budget of up to 60~m.s$^{-1}$ per year. Over a 5-year mission, this translates into a total budget of 300~m.s$^{-1}$.
In order to limit the dephasing with respect to the ground track to less than 1~km, these manoeuvers must occur, at most in the worst case scenario, every 48~h. This very high frequency of maneuvers leads to envision an autonomous, on-board system. The number of other satellites flying at such a low altitude  is limited therefore such a system would be satisfying even with debris catalog updated sparsely through mission-driven contacts with the ground.

\subsection{Spacecraft}

In this part, we present an analysis of the spacecraft architecture, including all the subsystems, which was carried out according to the mission scenario. 

\subsubsection{Impact of mission parameters on satellite architecture}

The described space geodesy mission induces significant constraints on satellite architecture. First, the measurements of gravity field require a low altitude polar orbit which implies a high atmospheric density and then a strong drag that have to be minimized. Second, the cold atom accelerometer must be placed at the center of mass of the satellite which must be kept at the same position throughout the life of the satellite. Finally, all dynamic disturbances must be minimized due to the high sensitivity of the cold atom accelerometer. These constraints have a strong impact on the spacecraft layout and then induce an interpenetration of the payload and the platform that have to be designed together.

\paragraph{Satellite shape:}

The mission constraints preclude having deployable solar panels, which would inevitably create vibrations on the instrument, so they must be on the spacecraft body. The mission also requires the satellite to be in constant geocentric pointing. As the orbital plane is inertial, the Sun's angle, with respect to this orbit, rotates of 360$^{\circ}$ per year. The size of the frontal area of the spacecraft is a direct driver of atmospheric drag. Thus, this surface needs to be minimized and the spacecraft cannot be arbitrarily high or wide. We looked for the best spacecraft shape to accommodate as much power as possible, limiting our search of spacecraft shape to right trapezoidal prism (as the GRACE spacecraft is), with solar panels set on the top face and the two side faces. For a constant frontal area, our numerical simulations have shown that the best shape to collect sunlight is actually to forego the top face entirely and have a right triangular prism. However, such a shape is not optimal for the equipment accommodation. Moreover, this optimal shape is a shallow maximum: still for constant frontal area, flattening the top of the triangle only affect the overall budget by a few percent. The shape of GRACE is actually a very good, albeit not perfect, optimum for maximizing the total amount of energy obtain from Sun.

\begin{figure}
\centering \resizebox{8cm}{!}{\includegraphics{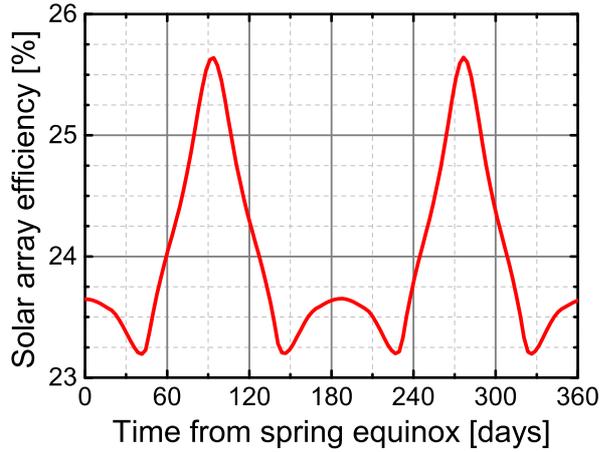}} 
%\centering \includegraphics[width=0.75\textwidth]{array.eps}
\caption{Solar array efficiency, averaged over 1 orbit on a full year. The initial mean local solar time of the ascending node is 12~h.}
\label{array}
\end{figure}

Over the course of the year, this amount of energy is approximately constant. The solar array efficiency averaged over an orbit is displayed on Figure~\ref{array}. The solar array efficiency is the ratio between the installed power and the average power obtained over one orbit. For instance, an efficiency of 25\% means that 4~kW of installed power would only generate 1~kW on average over the orbit. Although the RAAN plays a role in such a calculation, because we are at a near optimum configuration, its impact is not significant enough to alter the power budget calculation if changed. We see that, in our situation, the minimum efficiency, i.e. design point, of the solar array is of 23.2\% and is obtained neither at the equinoxes or solstices but in between. In other terms, for every 1~W of power required over the orbit, we should install 4.3~W of solar cells. 

\paragraph{Propulsion and attitude control subsystem:}

The last two constraints listed above have also an impact on the propulsion and attitude control subsystems. The center of mass position would be modified if using liquid propulsion. The liquid sloshing would also affect accelerometric measurements. So, for propulsion, only two possibilities remain: cold gas and electric propulsion. The tradeoff between these two options is realized for the attitude control and for the orbit control mode against the $\Delta V$ to maintain the orbit. For the attitude control subsystem, the choice of the actuators is driven by the third constraint. As micro vibration analysis was not carried out at this stage of the study, two options have been investigated, based on reaction wheels or propulsion. The Attitude Control Subsystem and associated modes are described in Appendix~\ref{app1}.

\subsubsection{Spacecraft layout}

\begin{figure}
\centering \resizebox{8cm}{!}{\includegraphics[angle=-90,origin=c]{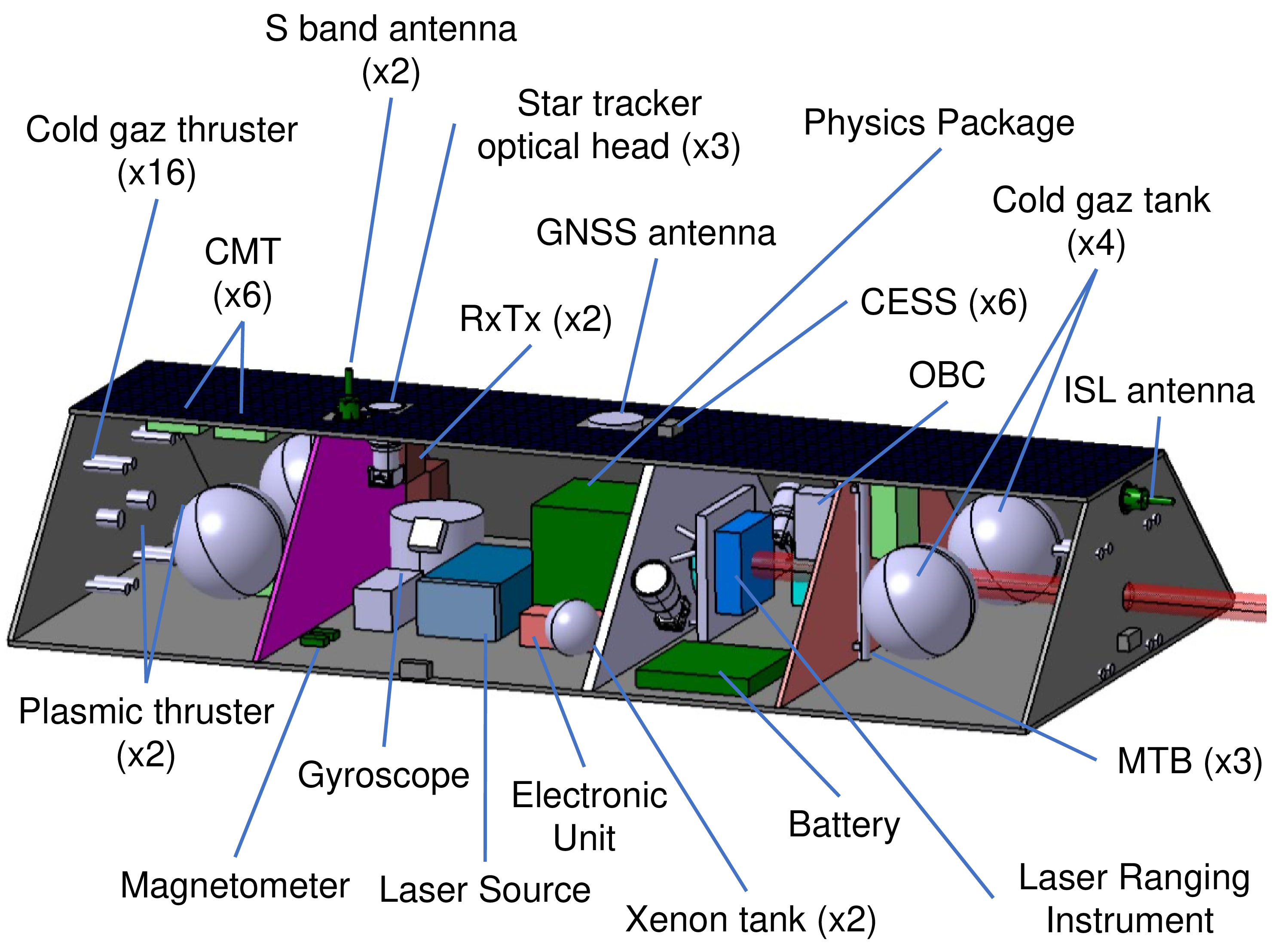}} 
%\centering
%\includegraphics[width=0.75\textwidth]{layout.eps}
\caption{View of the GRICE platform.}
\label{layout}
\end{figure}

The envisioned platform (Figure~\ref{layout}) provides all the necessary housekeeping functions to perform the mission: payload support, electrical power, thermal control, command, data handling and storage, attitude and orbit control. The platform structure has a trapezoidal shape. Its size is determined by the amount of energy necessary for the mission ($\approx$500~W in mission mode, $\approx$800~W in orbit control mode). Three faces support the solar cells coupled with a battery in order to provide the energy to the platform. A radiator is accommodated on the lower face for thermal control. The different parts of the payload are accommodated on the lower panel near the middle of the platform. The sensor part of the payload is placed at the satellite center of mass and supported by a transversal bench. The instrument, supported by a transversal bench, is installed on board the satellite so as to minimize off-centering between the position of the atomic cloud at the middle light pulse of the interferometer and the center of mass of the satellite. Six center of mass trim assembly (CMT, two per axis) are used to adjust center of mass as needed during the flight. The platform is based on 3-axes stabilization with nadir pointing. The attitude measurement is done by stellar sensor and gyroscopes. For the control, two options are identified: reaction wheels or propulsion (cold gaz) with magnetotorquers. At this stage of the study, the cold gaz is the reference solution with eigth thrusters placed on the velocity and anti-velocity faces of the platform associated with three magnetotorquers, one on each axe. For orbit control, two plasmic thrusters are accommodated on the anti-velocity face in order to cover the total $\Delta V$ budget of 300~m.s$^{-1}$. The data handling architecture is based on a central computer (OBC). The housekeeping TM/TC (telemetry/telecommand) and the payload telemetry is realized in S band. For the precise orbit determination, a bi-frequency GNSS receiver and laser reflector are also integrated on the platform. The mass of the satellite is around 900~kg.

\section{Gravity field restitution}
\label{sec:simulation}

\subsection{Simulation procedure}

Analytical simulations with dedicated methods can roughly express the expected contribution of a new mission design, but they are often optimistic. In the case of GRICE, many peculiarities intervene in the recovery of the gravity field such as the uncertainties in the ancillary computation models (tides in particular) or the rapid gravity changes in the Earth's surface layers (atmosphere-ocean interaction) that introduce aliasing phenomena.

This is why we conduct a numerical simulation to better examine the benefit of the GRICE concept in a supposedly realistic environment projected in 10 to 15 years time. In the context of this future mission, the noises impacting the different models are then supposed to be a fraction of the present day model differences. Moreover, the specified instrumental noises are introduced in the numerical simulation.

The goal is to evaluate the gravity gradient approach which involves two kinds of data: absolute acceleration measurements performed by the cold atoms accelerometers coupled with inter-satellite acceleration measurement obtained from the double differentiation of the inter-satellite laser link range measurement.

The numerical simulation relies on two sets of data. First a set of "true" data including gravitational accelerations computed from un-perturbed gravity models, non-gravitational accelerations computed from drag, solar and terrestrial radiations in the visible spectrum as well as Earth's infrared radiations and inter-satellite distance measurements computed from orbit propagated with the set of "true" models. Second, a set of "perturbed" data including noisy a priori gravity models, noisy accelerometer and inter-satellites measurements.

\subsection{Orbit scenario}

The mission orbit scenario has been presented before. It comes under several geometrical and dynamical considerations such as a global Earth's coverage, a monthly orbit repeatability and an optimal satellite inter-range for gravity field sensitivity. In the frame of the numerical simulation, the inclination is chosen to be 89$^{\circ}$ in order to keep a small crossing angle of the ground tracks, without losing any signal at the poles. The choice of flying altitude was not challenged during the numerical simulations that were carried out afterwards to assess more precisely the performances of GRICE with regards to the gravity field.

\subsection{Algorithms}

In the case of the GRACE mission, gravity field spherical harmonic coefficients~\cite{bettadpur2018,tapley2004_2,flechtner2010,dahle2014} are usually adjusted numerically from the range measurement between satellite B and satellite A provided by the K-band ranging (KBR) instrument. The non-gravitational accelerations measured by the accelerometers on board each satellite are added to the ones of the gravitational models for performing a double integration (using the Cowell method for instance) which provides orbit positions of both satellites. The projected difference on the line-of-sight ($L$) can then be compared with the KBR observations taking into account all geometric or physical corrections:

\begin{equation}
L= ||\mathbf{r_B} - \mathbf{r_A} ||,
\end{equation}

with $\mathbf{r_{A,B}}$ being the geocentric distance of the satellites A and B. But the most general approach is to use the time derivative ($dL/dt$), called K-band range-rate (KBRR), in order to reduce observation biases and to increase the gravitational orbit perturbation sensitivity:

\begin{equation}
\frac{dL}{dt}=\frac{(\mathbf{v_B}-\mathbf{v_A}) \cdot (\mathbf{r_B}-\mathbf{r_A})}{L},
\end{equation}

with $\mathbf{v_{A,B}}$ being the satellite velocities. As one would be able to derive twice the inter-satellite distance with enough precision,it is possible to work at the acceleration level considering bias-free accelerometers to form directly the gravity gradient observable. The range acceleration is then:

\begin{equation}
\frac{d^2L}{dt^2}=\frac{(\mathbf{a_B}-\mathbf{a_A} ) \cdot (\mathbf{r_B}-\mathbf{r_A} )+ || \mathbf{v_B} - \mathbf{v_A} ||^2 - \frac{dL}{dt}^2}{L}, 
\end{equation}

 with $\mathbf{a_{A,B}}$ being the satellite accelerations. This is what we call the gradient approach.

\begin{figure}
\centering \resizebox{8.5cm}{!}{\includegraphics[angle=-90,origin=c]{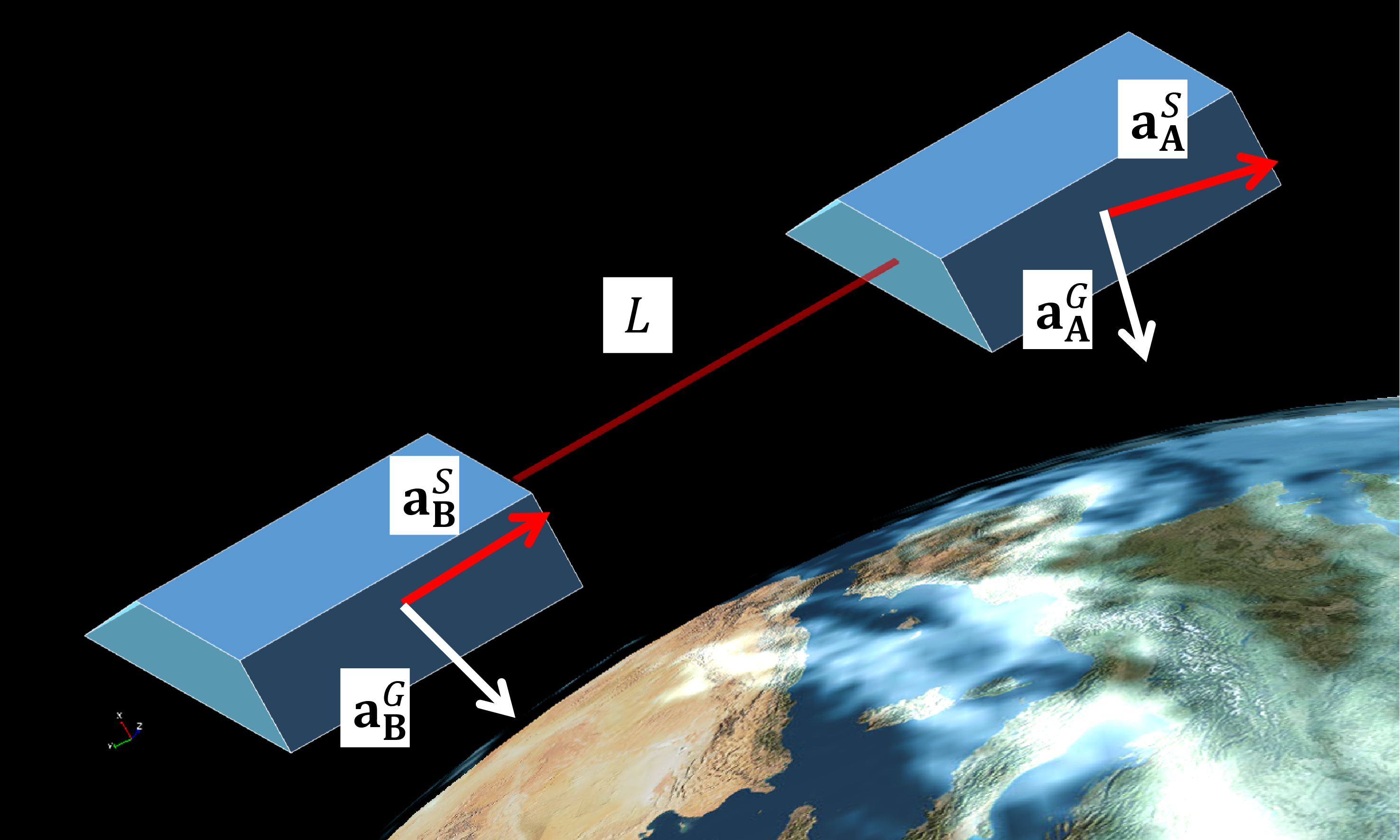}} 
%\centering \includegraphics[width=0.75\textwidth]{diagram.eps}
\caption{Illustration of the accelerations acting on the satellites. The white arrows labeled $\mathbf{a_{A,B}}^G$ represent the gravitational accelerations acting on the satellites. The red arrows labeled $\mathbf{a_{A,B}}^S$ represent the surface accelerations acting on the satellites.}
\label{sat_acceleration}
\end{figure}

In fact, both satellites experience gravitational ($\mathbf{a}^G$) and surface ($\mathbf{a}^S$) accelerations which can be decomposed as follows (see Figure~\ref{sat_acceleration}):

\begin{equation}
\mathbf{a_B} -\mathbf{a_A}=(\mathbf{a_B} - \mathbf{a_A} )^G - (\mathbf{a_B} - \mathbf{a_A} )^S
\end{equation}
 
The projection of gravity acceleration differences on the line-of-sight is equivalent to the gradient of the gravitational potential at the middle of the inter-satellite distance. If one isolates the modelled part of gravity ($\mathbf{a}^{Gm}$) from the un-modelled part (hydrology, $\mathbf{a}^{Gh}$), it yields:

\begin{equation}
\begin{split}
\frac{(\mathbf{a_B} - \mathbf{a_A} )^{Gh} \cdot ( \mathbf{r_B}-\mathbf{r_A} )}{L}=\left(\frac{d^2L}{dt^2}\right)^2\\
+\frac{\frac{dL}{dt}^2 - || \mathbf{v_B}-\mathbf{v_A} ||^2}{L}\\
-\frac{(\mathbf{a_B} - \mathbf{a_A})^S \cdot (\mathbf{r_B}-\mathbf{r_A} )}{L}\\
-\frac{(\mathbf{a_B} - \mathbf{a_A} )^{Gm} \cdot (\mathbf{r_B}-\mathbf{r_A})}{L}
\end{split}
\end{equation}

where the different terms represent the second derivative of the inter-satellite distance measurement, the velocity correction, the accelerometer measurements and the modelled gravity,  respectively.
 
The gradient approach is promising if the numerical derivation of the inter-satellite distance can be maintained at a low level of noise. In this respect the noise of GRACE KBR measurements appears to be too high, which explains that the classical KBRR approach is generally used~\cite{goswami2018}. This is no more the case for laser-range interferometry (LRI) which is planned to be used for GRICE, with a noise supposed to be up to 100 times lower. Hence the GRICE simulations are performed with the gradient approach over 2 years of data (2006-2007 environmental conditions). Monthly gravity field models are computed and compared to a priori hydrology models as defined in the following. However, a comparison with the range-rate approach is worked out as well in order to assess the benefits of the gradient approach.

\subsection{Model noise}

\begin{figure}
\centering \resizebox{8.5cm}{!}{\includegraphics{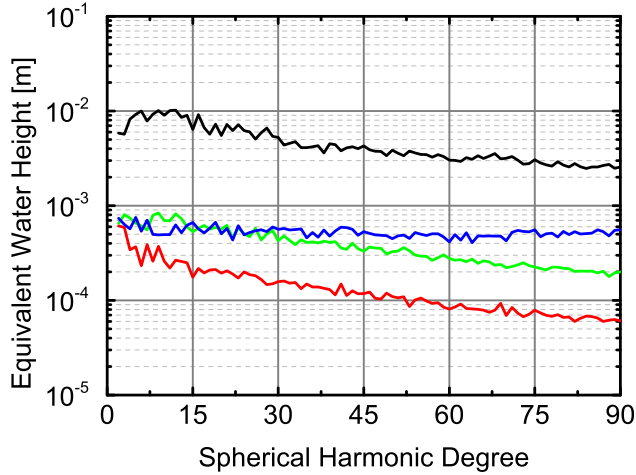}} 
%\centering \includegraphics[width=0.75\textwidth]{diagram.eps}
\caption{Spectrum of hydrology signal (black) considered at the input of the simulation for a given date of the simulated period. We report on the same graph the error spectra considered for each input model (atmosphere in red, ocean response in green, ocean tides in blue).}
\label{model_errors}
\end{figure}

The simulation of model noise is built from the model differences for ocean tides, atmosphere pressure, ocean response to surface pressure and wind stress, hydrology (Table~\ref{tab7}). The full difference of nowadays models~\cite{mayer2018} appears as too large for a mission to be launched in a decade or more, so we consider only 20\% of the model differences. Moreover, we amplify by a factor 2 the hydrology signal to be detected because the current hydrological models (GLDAS, WGHM…) seem to be too weak in amplitude compared to GRACE's solutions~\cite{landerer2012}. An example of comparison (in terms of spherical harmonic expansion expressed in equivalent water height), at one particular date, between the hydrology signal and the model errors is shown in Figure~\ref{model_errors}.

\begin{table*}
% table caption is above the table
\caption{Models used in data simulation (A) and for the gravity field adjustment (B).}
\label{tab7}       % Give a unique label
% For LaTeX tables use
\centering
\begin{tabular}{llll}
\hline\noalign{\smallskip}
 & "True" models used to &Alternative models (B)&A priori models used in the \\
 & generate the "true" orbit (A) & & adjustment procedure\\
\noalign{\smallskip}\hline\noalign{\smallskip}
Static gravity field&	CNES-GRGS-RL03.v2&	ITSG-Grace2014&	A+20\%(B-A) for degrees \\
 & & & 91 to 120 (omission error)\\
Ocean tides&	FES2014 (20 waves)&	GOT4.8 (20 waves)&	A+20\%(B-A)\\
Atmospheric pressure&	ECMWF/6h&MERRA2/6h&	A+20\%(B-A)\\
Ocean response&	T-UGOm/6h&	OMCT/6h&	A+20\%(B-A)\\
Hydrology&	GLDAS/6h signal completed&	none&	2 $\times$ A\\
 & by GRACE data at high latitudes & & \\
\noalign{\smallskip}\hline
\end{tabular}
\end{table*}

\subsection{Instrumental noise}

Instrumental noise is directly applied on the “true” measurement files. As presented in the part~\ref{sec:payload}, we consider that the two types of instruments involved in the measurement (i.e. Atom accelerometers and LRI) are white noise limited. The gravity recovery algorithm used for these numerical simulations needs input data with a sampling time of 5~s. As the cycling time consider for our instruments is about 2~s, we consider the following noise levels:

\begin{itemize}
\item $2 \times 10^{-8}$ m at 1 $\sigma$ for inter-satellites range measurements. Range-rate and accelerations measurements are derived using a polynomial approach over 100 s in sliding window. 
\item $3 \times 10^{-10}$ m.s$^{-2}$ at 1 $\sigma$ for accelerometer measurements. No calibration parameters are introduced considering the absolute nature of atomic accelerometers. Indeed, atom accelerometers are known for having an excellent accuracy (i.e. stable and well characterized bias) as shown in~\cite{karcher2018,louchet2011}. Therefore, neither systematics nor drifts are introduced in the simulation.
\end{itemize}

\begin{figure}
\centering \resizebox{8.5cm}{!}{\includegraphics{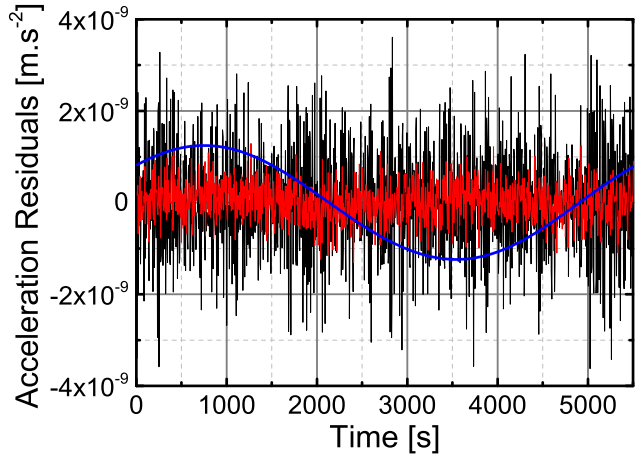}} 
%\centering \includegraphics[width=0.75\textwidth]{diagram.eps}
\caption{Acceleration residuals over one satellite revolution considering the 3 kinds of data noise. LRI noise (black), accelerometer noise (red), ephemeris noise (blue). The 1 cm random ephemeris noise generates mostly a once-per-revolution residual signal in acceleration at some $10^{-9}$ m.s$^{-2}$.}
\label{acceleration_residuals}
\end{figure}

Although the gravity field recovery is done with only the knowledge of accelerometer and inter-satellites range data, it is useful to constraint the absolute 3D-position of GRICE orbit at some centimeter level (i.e. equivalent to GNSS positioning). In fact, we introduce a priori ephemeris data with a random noise of 1~cm (at 1 $\sigma$) per coordinate. 
The double derivation of the range introduces of course some additional noise which is take into account in the simulation. This noise is compared to the accelerometer noise on Figure~\ref{acceleration_residuals}.
Systematic noises are then automatically introduced through the imperfections of the variable gravity models.

\subsection{Methodology}

In this work, 730 one-day arcs are computed with the nominal models to generate the “true” measurement files. Once the measurements have been noised and derived, the same 1-day arcs are computed using the alternative models with the following partial derivatives:
\begin{itemize}
\item Initial arc parameters;
\item Empirical once per revolution periodic coefficients in each direction (radial, tangential and normal) per day;
\item Gravity field spherical harmonic coefficients from degree 2 to 90.
\end{itemize}

Normal equations are formed per day and stacked per month. The 24 monthly normal equations are then solved using two different inversion methods. The first one is the Cholesky decomposition without any constraint. The second one is the Singular-Value Decomposition method (SVD) eliminating the smallest non-significant eigenvalues in order to stabilize the solutions.

\subsection{Results}

\begin{figure}
\centering \resizebox{8.5cm}{!}{\includegraphics{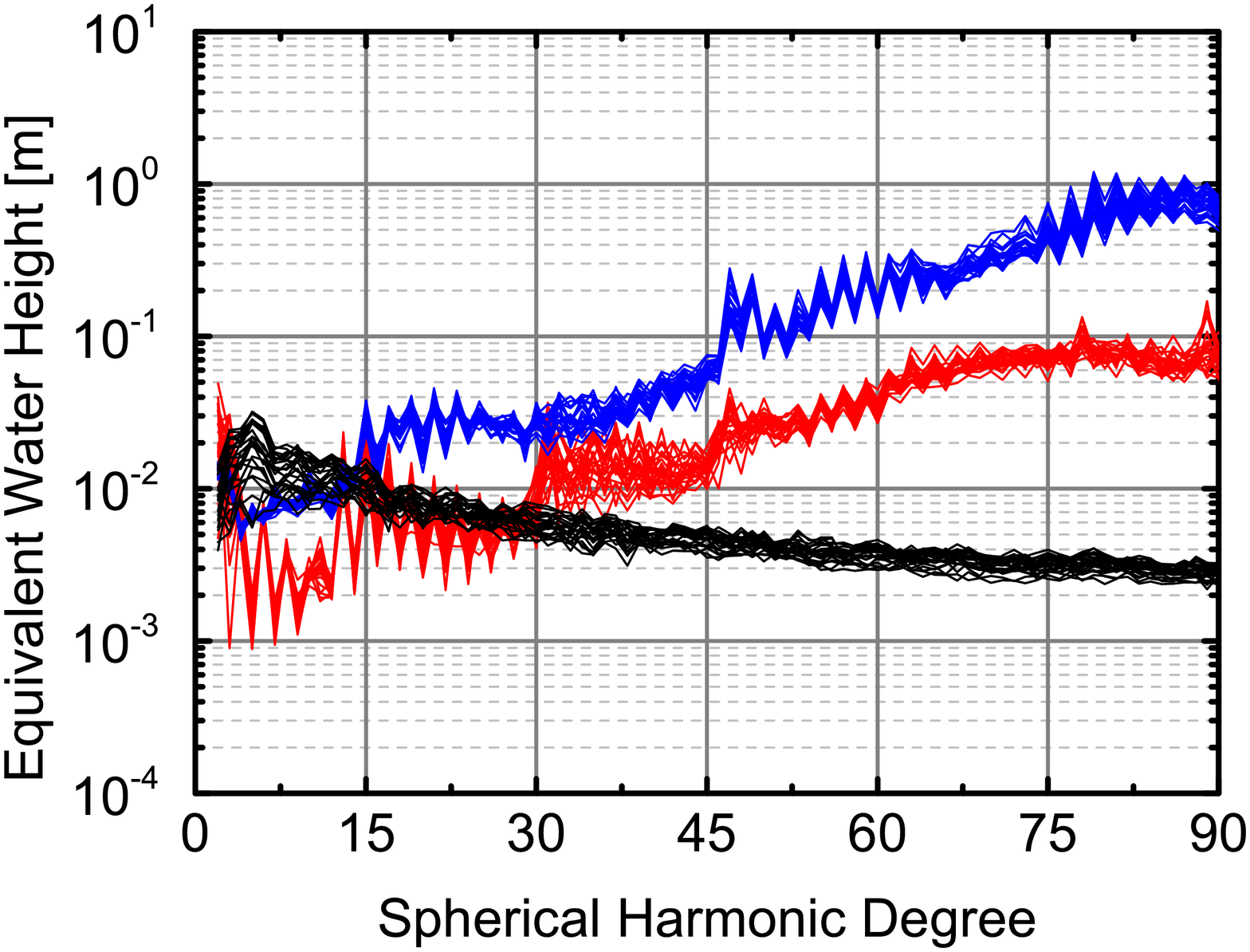}} 
\centering \resizebox{8.5cm}{!}{\includegraphics{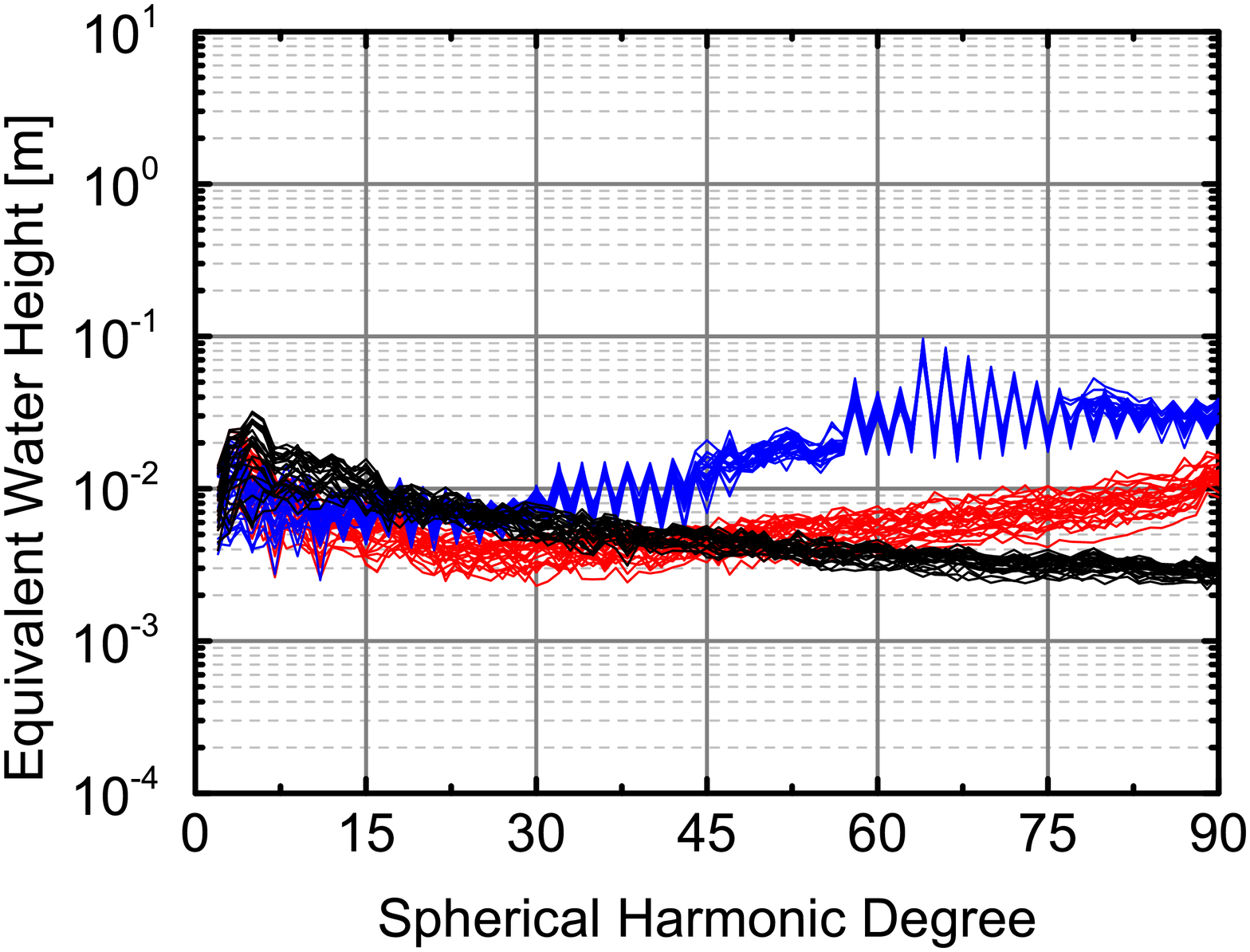}}
%\centering \includegraphics[width=0.75\textwidth]{diagram.eps}
\caption{Spectra of gravity field recovery in equivalent water height from Cholesky (top) and SVD (bottom) inversions of LRI acceleration (in red) and LRI range-rate data (in blue). The hydrology signal to be recovered is in black.}
\label{spectra}
\end{figure}

The inherent problem of the spherical harmonic representation in orbital adjustment is the loss of sensitivity at the higher degrees. A global inversion through the Cholesky decomposition for instance is not stable enough and requires some stabilization process afterwards. We prefer to use the SVD method which produces cleaner solutions provided that we truncate the number of eigenvalues to the most significant ones. The level of truncation is empirically fixed according to noise and track artifact apparition. The number of linear combinations of Stokes coefficients that are solved is much higher for the acceleration approach, for which 75\% of eigenvalues are retained, than for the range-rate approach where, on the opposite, 75\% of eigenvalues are eliminated. Figures~\ref{spectra} show the spectra of the 24 monthly gravity field solutions (in equivalent water height) for both cases, Cholesky and SVD, obtained from acceleration data (in red) compared to the ones obtained from range-rate data (in green). It appears that the SVD method reduces the noise up to one order of magnitude from degrees 50 to 90 and that the acceleration method gives systematically less noisy results. 

\begin{figure}
\centering \resizebox{8.5cm}{!}{\includegraphics{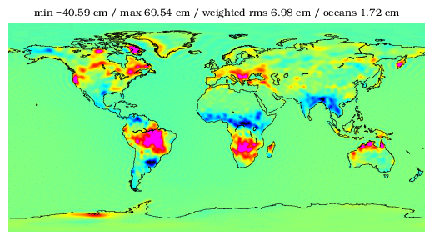}} 
\centering \resizebox{8.5cm}{!}{\includegraphics{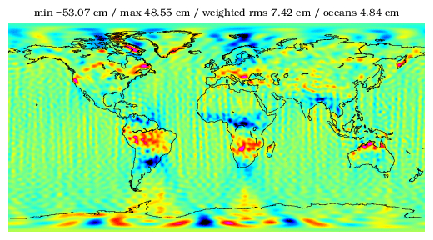}}
\centering \resizebox{8.5cm}{!}{\includegraphics{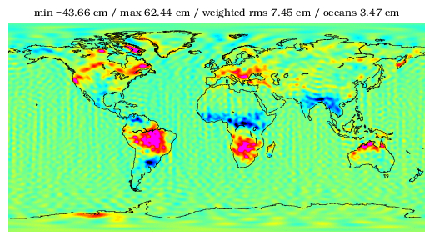}}
\centering \resizebox{8.5cm}{!}{\includegraphics{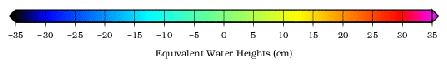}}
%\centering \includegraphics[width=0.75\textwidth]{diagram.eps}
\caption{Hydrology anomaly (signal to be recovered) for March 2006 (top), adjusted signal from degree 2 to 90 (middle) and adjusted signal from degree 21 to 90 when introducing the monthly mean of a priori hydrology coefficients up to degree 20 (bottom).}
\label{map}
\end{figure}

However, some artifacts remain in the acceleration solutions as shown in Figure~\ref{map}(top) (modeled hydrology signal for March 2006) and Figure~\ref{map}(middle) (recovered signal after SVD inversion). In fact, it appears that acceleration measurements are less sensitive to very low spherical harmonic degrees. This fact tends to attenuate the long wavelength signal in the solution and generates some artifacts mainly at the poles. If one assumes that the degree 2 to 20 are known by any other type of data (like ranging data), the acceleration method allows to recover almost perfectly the hydrology signal originally introduced (Figure~\ref{map}(bottom)).

\begin{figure}
\centering \resizebox{7cm}{!}{\includegraphics{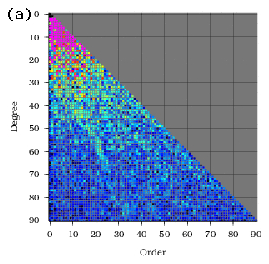}} 
\centering \resizebox{7cm}{!}{\includegraphics{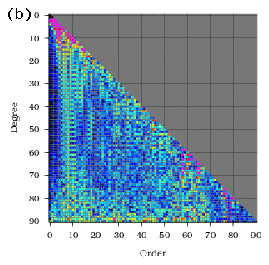}}
\centering \resizebox{7cm}{!}{\includegraphics{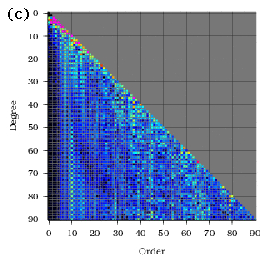}}
\centering \resizebox{7cm}{!}{\includegraphics{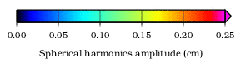}}
%\centering \includegraphics[width=0.75\textwidth]{diagram.eps}
\caption{Hydrological signal to be recovered (a) with respect to spectral degrees between 0 and 90 for March 2006, differences with the adjusted coefficients (b) and with the solution without any noise (c).}
\label{harmonics}
\end{figure}

Another way to assess the gravity field solutions is to look at the agreement in terms of spherical harmonic coefficients. The bloc diagram of coefficient amplitudes given in terms of equivalent water height compared to the hydrology signal to be recovered is a good visual indicator of the quality of the adjustments. Thus, the Figure~\ref{harmonics}a shows coefficients amplitudes of the modeled hydrology signal for March 2006 corresponding to Figure~\ref{map}(top). The Figure~\ref{harmonics}b presents the difference obtained after adjustment. It is clear that the low sectorial and near-sectorial coefficients (under degree 20) are not well adjusted and that their errors propagate around order 9. In fact, this is due to the technique itself as it proved by a zero test (Figure~\ref{harmonics}c) that consists of adjusting the hydrology signal without any simulated noise either from models or from the instruments.

Compared to current GRACE/GRACE-FO solutions, the combination of a low flying altitude for GRICE and of the gravity gradient approach allows a reduction of amplitude of the degree-error curves in the higher degrees and hence a better spatial resolution. This was verified by comparing the red curve of Figure~\ref{spectra} (top), with the formal error curves of the ITSG-Grace2018 solution from the Institut fur Geodesie (IfG) of Graz Technical University (TUGRAZ), for the years 2006-2007. We selected this solution because the computation of the formal error for these models is done with extreme care and represents a realistic estimate of the true error of the GRACE solutions~\cite{mayer2018}.
While the error curves of ITSG-Grace2018 align almost on a straight line (in a linear-log plot) between 1 mm EWH at degree 3 and 30-40~cm EWH at degree 90, with an intermediate point at 2~cm EWH at degree 45, it can be seen on Figure~\ref{spectra}, that the error curves of GRICE display a lower slope after degree 45, ending up at 7-8~cm EWH at degree 90. As a result, GRICE allows a better determination, by a factor $\sim$5, of the shorter wavelengths of the gravity field than GRACE/GRACE-FO.

\subsection{Comparison between acceleration and range-rate methods}
\label{comparison}

All computation presented here have been done without empirical periodic parameter adjustment. In that case, the acceleration method applied from degree 2 to 90 performs much better than the range-rate method as described in Table~\ref{tab8} (up to almost 5 times). However, introducing once per revolution parameters per day in the adjustment reduces the difference to 8\% between both approaches. The results are approaching the level when no instrumental or model errors are introduced.

\begin{table*}
% table caption is above the table
\caption{Impact of adding once per revolution acceleration parameters in radial/tangential/normal directions on range-rate and acceleration residuals worldwide, expressed in cm of EWH for Spherical Harmonics (SH) 2 to 90.}
\label{tab8}       % Give a unique label
% For LaTeX tables use
\centering
\begin{tabular}{lll}
\hline\noalign{\smallskip}
Global RMS error (cm of EWH) & Range-rate (SH 2 - 90) & Acceleration (SH 2 - 90)\\
\noalign{\smallskip}\hline\noalign{\smallskip}
Without any additional parameters & 21.26 & 4.35 \\
With once per rev. parameters & 3.98 & 3.63 \\
Without any instrumental or model noises & 3.40 & 3.12 \\
\noalign{\smallskip}\hline
\end{tabular}
\end{table*}

Nevertheless, the improvement is not global. As depicted before, the acceleration method tones down the low degrees compared to the range-rate method usually applied to the GRACE KBR data. That is indicated in Table~\ref{tab9} when splitting the harmonic decomposition in 3 brackets of low degrees (2 to 20), medium degrees (21 to 60) and higher degrees (61 to 90).

\begin{table*}
% table caption is above the table
\caption{RMS errors by brackets of spherical harmonic expansion over oceans, continents and global for both adjustment types in terms of range-rate and acceleration measurements.}
\label{tab9}       % Give a unique label
% For LaTeX tables use
\centering
\begin{tabular}{lclclclclclcl}
\hline\noalign{\smallskip}
SH degrees & \multicolumn{2}{c}{Oceans (cm)} &  \multicolumn{2}{c}{Continents (cm)} & \multicolumn{2}{c}{Global (cm)}\\
%\noalign{\smallskip}\hline\noalign{\smallskip}
\noalign{\smallskip}\cline{2-7}\noalign{\smallskip}
& Range-rate & Acceleration & Range-rate & Acceleration & Range-rate & Acceleration\\
\noalign{\smallskip}\hline\noalign{\smallskip}

  2 - 20 & 1.36 & 1.34 & 2.50 & 2.90 & 1.86 & 2.05\\
21 - 60 & 1.79 & 1.22 & 2.71 & 2.16 & 2.17 & 1.63\\
61 - 90 & 2.12 & 2.16 & 3.61 & 3.02 & 2.76 & 2.51\\
global: 2 - 90 & 3.00 & 2.79 & 5.26 & 4.84 & 3.98 & 3.63\\

\noalign{\smallskip}\hline
\end{tabular}
\end{table*}

\subsection{Contribution of atom accelerometers to the mission performances}

As presented in the section~\ref{sec:payload}, the use of cold atom accelerometers is required for retrieving a long baseline gradiometric measurement. Indeed, two main issues can arise when coupling the instruments to retrieve the gravity gradient measurement. The first one is the imperfect rejection of common mode noises and the second is the knowledge of the bias affecting each accelerometer and their drift over time. In this work, we highlight two main aspects of this mission concept that ease the coupling between these instruments. First, the use of common mirrors between the laser link and the atom accelerometers ensures a high rejection of common mode noises. Second, atom accelerometers are known to have an excellent accuracy (i.e. stable and well characterized bias). This characteristic eases the combination of the two measurements. Therefore, the gradient of acceleration will not be affected by any significant bias, specially over long timescales. Thus, the performances of the atom accelerometers, which provide a low-bias and low-drift measurement, is the keystone of this gradient approach.

In order to point out the interest of atom accelerometers in the specific case of this mission scenario, we run a simulation to determine the performances which would be achieved by a drifting accelerometer in the same mission scenario. In particular, we focus on the impact of long term stability, in the 10$^{-6}$~Hz to 10$^{-5}$~Hz frequency band, on the gravity restitution. Indeed, accelerometers used in space missions generally exhibit an increase of the noise at low frequency, induced by thermal effects on the system~\cite{berge2013}, that could be avoided using atom accelerometers~\cite{lautier2014}. The methodology used for this simulation consists in adding drifts in the instrumental noise considered at the input of the numerical calculation. This approach is a simple method for modeling the behavior of these instruments in the spectral band of interest. Therefore, this simulation enables to retrieve the gravity field in the same conditions for both a drifting accelerometer and quantum technologies.

In order to simulate the performances of a drifting accelerometer, a drift and a periodic signal have been introduced on the accelerometric measurements (different on each satellite). When considered on a few days timescale, a typical drift of $10^{-9}$~m.s$^{-2}$ per day has been added. We also introduce a periodic signal at 10800 seconds with an amplitude of about $3 \times 10^{-10}$~m.s$^{-2}$, corresponding to a typical perturbation related to the orbital period.

All computations were carried out without any empirical periodic parameter adjustment in the gravity field restitution. This approach enables to compare the technologies with the same level of signal post-processing. Indeed, the adjustment of empirical parameters, which generally limit the effect of instrumental drifts, is also likely to affect the gravitational signal.

\begin{figure}
\centering \resizebox{12cm}{!}{\includegraphics{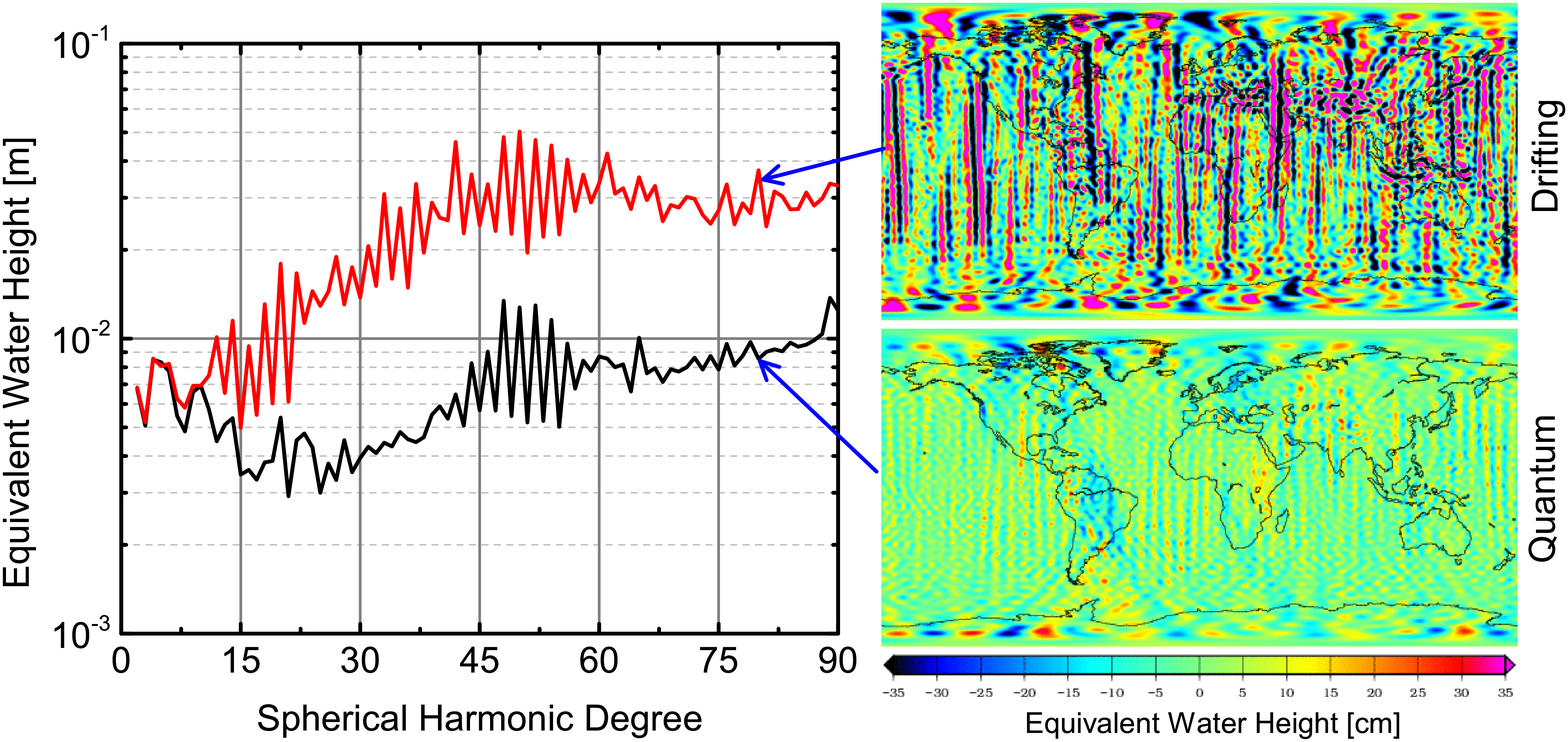}}
%\centering \includegraphics[width=0.75\textwidth]{diagram.eps}
\caption{Spectra of gravity field recovery in equivalent water height considering quantum (black) or drifting (red) accelerometers. The computations were carried out without any empirical periodic parameter adjustment in the gravity field restitution.}
\label{quantum_vs_classical}
\end{figure}

The Figure~\ref{quantum_vs_classical} exhibits the error spectra of the gravity field solution in equivalent water height for both quantum and drifting accelerometers. We also report the gravity maps related to each case. It appears that the quantum accelerometers enable to reduce the noise up to a factor of 5 from degree 50 with respect to the drifting accelerometers. The global RMS error drop from 23.64 cm of EWH for the drifting technology to 6.72 cm of EWH for the cold atomic sensor, improving the quality of the gravity map restitution.

Further simulation also shows that the gap between these two results is reduced by adjusting empirical parameters in the gravity field restitution. However, this study highlights that the long-term stability of accelerometers is an important parameter for space geodesy missions. Moreover, atom accelerometers need no calibration of the scale factor which is perfectly known and stable over time. Thus, these characteristics would enable to get rid of empirical parameters during data processing and thus potentially improve the quality of the gravity signal restitution.

\section{Conclusion}

We carried out an assessment study of a space mission for gravity field mapping involving innovative atomic technologies. The related mission scenario implies a constellation of two satellites each equipped with a cold atom accelerometer placed at their center of mass. A laser link measures the distance between the two satellites and couples these two instruments in order to produce a correlated differential measurement. This instrument enables to retrieve a long baseline gradiometric measurement along the track of the two satellites which is likely to increase the resolution of the high harmonics of the Earth's gravity field. The main parameters of the instruments have been estimated in order to determine the final performances of the payload. To ensure an optimal temporal resolution, the mission's orbit has been optimized to provide a consistent coverage over different time scales. Finally, a preliminary design of the platform has been realized. This design integrates all the subsystems which enable the satellite to operate at low orbit during 5 years and demonstrates the technical feasibility of this concept. 

Further work has been performed in order to simulate the gravity field recovery. For this, a specific data processing method has been developed taking advantage of the correlation between the two atomic accelerometers by a gradient approach. The simulations have shown that, with the given level of instrument specifications and with the chosen model type errors, this method would be able to give its best performance in terms of monthly gravity fields recovery under 1000~km resolution. In the resolution band between 1000 and 222~km, the improvement of the GRICE gradient approach over the traditional range-rate approach is globally in the order of 10 to 25\%. An alternative method would be necessary for the lowest degrees either in iteration or in combination with a KBRR-type determination, for an optimal recovery of the time variable gravity signal. The use of GNSS-type inter-satellite range data for determining the low degree coefficients might be the best complementary approach.

This study highlights the potential of atom accelerometers in space for Earth's science applications. In particular, it shows that long term stability and accuracy specific to these instruments could improve the determination of the Earth's gravity field. Such improvement of the gravity map resolution would pave the way for new fields of applications in hydrology, glaciology, oceanography and internal geophysics. Indeed, it would enable a better monitoring of mass displacements in both internal structure and external fluid layers and then contribute to a better understanding of the Earth's system and, in particular, its global climate change.

To a greater extent, the mastery of atom accelerometers in space will become a determining scientific and technical stakes in the next decades. In this context, other work will be conducted at CNES in order to study a pathfinder mission for this type of instrument. This new proposition of mission, called CARIOQA (Cold Atom Rubidium Interferometer in Orbit for Quantum Accelerometry) will demonstrate the utilization of a quantum inertial sensor in space paving the way for future space applications in Earth's science and fundamental physics.

%
% For two-column wide figures use
%\begin{figure*}
% Use the relevant command to insert your figure file.
% For example, with the graphicx package use
 % \includegraphics[width=0.75\textwidth]{example.eps}
% figure caption is below the figure
%\caption{Please write your figure caption here}
%\label{fig:2}       % Give a unique label
%\end{figure*}
%

\begin{acknowledgements}
 We would like to thank I. Panet for providing insightful information about application of mission data in the frame of Earth's science, and for careful reading of the paper. We also thank G. Ramillien, M. Diament, S. Bonvalot, B. Foulon, A. Gauguet and I. Petitbon for fruitful discussions in the frame of the GRICE mission group.
%If you'd like to thank anyone, place your comments here
%and remove the percent signs.
\end{acknowledgements}

\section*{Author contributions} TL and CF conducted the study and wrote the paper; MM defined the general scientific scope of the study; FPDS, PB and BB determined the performances of the atom accelerometers; ST performed the optimization of orbital parameters; SD and AP designed the satellite platform; RB, JML and SB conducted the simulations of Earth’s gravity field recovery.

\section*{Data Availability Statement} All data sets are available upon reasonable request from the authors.

% Authors must disclose all relationships or interests that 
% could have direct or potential influence or impart bias on 
% the work: 
%
% \section*{Conflict of interest}
%
% The authors declare that they have no conflict of interest.

\appendix

\section{Attitude Control Subsystem} 
\label{app1}

The mission definition and constraints have a strong impact on the Attitude Control System (ACS) architecture. The nadir pointing shall be maintained within 4~mrad and 50~$\mu$rad/s. This nonetheless ensures a correct alignment of the payloads of the two spacecraft with respect to one another, but it also minimizes the impact of air drag (and thus increases the mission lifetime). ACS shall also provide the estimated attitude to the instrument with a knowledge error lower than 0.1~mrad. Such requirements are not particularly stringent, but due to the low altitude of the orbit and the elongated shape of the spacecraft, the impact of orbit related perturbations has to be carefully considered in the ACS design. Indeed, with such a long shape, the air drag and gravity gradient torques are strongly dependent of the pointing error. The spacecraft layout also plays an important part in the air drag torque: any offset between the center of mass and the aerodynamic center increases the aerodynamic torque linearly. Moreover, in addition to the attitude control during the measurements sessions, ACS shall ensure a correct pointing during the orbit correction manoeuvers and in case of failure. To reply to all these considerations, ACS is defined to provide three-axis stabilized Earth-pointing attitude control during all mission phases.  Thus, ACS architecture is composed of a set of equipment and on-board software used within the control closed loop. Sensors are used to measure the attitude and actuators to correct it. Estimation and attitude control algorithms are implemented on-board to compute the estimated attitude and the control law. Three ACS modes are defined depending on the phase of the mission described below:

\paragraph{Mission mode:}

This is the mode during which the mission is carried out. In this mode, the control loop uses measurements from a 3-head Star-Tracker (STR) and a 3-axis optical gyroscope to estimate the attitude. As the orbital plane is inertial, a 3-head STR is needed: the Satellite-Sun angle varies along the mission (which induces successive dazzling of each panel of the satellite), the heads are thus set on the lateral and upper panels to ensure at least one available STR. Both measurements are hybridized in a gyro-stellar filter to compute the attitude. 

Concerning the actuators, in order to minimize the perturbations toward the payload, two options are considered to realize the commanded torques: a set of 4 reaction wheels in pyramidal configuration (to ensure the torque capacity and a redundancy in case of failure) or a cold gas propulsion subsystem (CGPS). The CGPS configuration is composed of $4 \times 2$ thrusters on the front and back sides of the satellite. Magneto-torquer bars (MTBs) are also used in both options: either to unload the wheels, or to help the CGPS to compensate for the Earth's magnetic torque.  

\paragraph{Safehold mode:}

This mode is enabled just after separation to stabilize the satellite (inertial acquisition) and in case of failure detection or attitude loss. The satellite remains in geocentric pointing to minimize air drag perturbations, but with relaxed pointing and stability performances. The attitude estimation uses a set of Sun sensors and magnetometers. Six Coarse Earth Sun Sensors (one head on each of the six sides of the satellite) are used to give omnidirectional and coarse attitude estimation. In addition, 2 fine Sun sensors, located on the lateral sides are used to provide a precise attitude once the satellite is stabilized. Two magnetometers are used to estimate the angular rate. The 3-axis stabilization is performed by MTBs.

\paragraph{Orbit Control mode:}

During the orbit control maneuvers, the thrust direction has to be maintained along track to counter air drag effect. Consequently, the satellite has to remain in nadir pointing. The orbit control propulsion system is not suited to realize attitude control by off-modulation. ACS uses the same sensors and actuators as the Mission mode (star tracker and gyroscope as sensors, and reaction wheels or CGPS as actuators), whilst the thrust is performed by electric propulsion. 

In this early phase of the project, particular attention is paid to the dimensioning of the actuators (MTBs, reaction wheels, CGPS). Indeed, the satellite layout is strongly dependent of the size and mass of the actuators. The dimensioning process consist in checking that the capacities of the chosen actuators are compliant with the main perturbations encountered for each mode. 

In Safehold mode, MTBs have to deal with potential huge pointing errors and thus high aerodynamic and gravity gradient torques. Nevertheless, in Mission and Orbit Control modes, the main constraint is the unload duration of the reaction wheels that has to be lower than a quarter orbital period.

For the reaction wheels and the CGPS, the dimensioning constraint comes from the orbit control maneuvers. Indeed, the thrust torque is directly linked to the offset between the center of mass and the aerodynamic center of the platform: with a 300~m/s $\Delta V$ budget over 5 years shared with manoeuvers every 48~h, each 2h-spread tangential thrust generates a constant 2.5~mN.m torque in the case of 50~mm offset (in pitch/yaw). Moreover, concerning the reaction wheels option, an unloading by MTBs during the manoeuver is necessary to be compatible with the wheels’ angular momentum. Concerning the CGPS option it is also important for the satellite layout to estimate the gas consumption during both modes: with the considered hypotheses (5-year mission, 50~mm offset, 55~s Isp, 1000~kg satellite mass) the foreseen gas consumption is 35~kg in Mission mode and 15~kg in Orbit Control mode. This consumption is a major factor in the trade-off between the two options.

The trade-off between reaction wheels and cold gas propulsion system is not completely done at this stage of the project. Both solutions are compatible with the mission needs. Reaction wheels are often used as main actuator in ACS, mainly because they don’t need any propellant. They also provide the satellite attitude control system with a high torque capacity. However, they induce micro-vibrations which may be incompatible with atomic accelerometer measurements (the effect of micro-vibrations on the instrument have not been characterized yet). On the other hand, CGPS certainly induces a gas consumption which has to be taken into account in the satellite layout, but the feedback of high precision missions such as Microscope~\cite{delavault2019} or GAIA showed a very satisfactory level of performance well suited to GRICE needs. At this stage of the study, the CGPS is then the reference solution for the satellite layout. During this mode, no scientific measurements will be realized.

% BibTeX users please use one of
%\bibliographystyle{spbasic}      % basic style, author-year citations
%\bibliographystyle{spmpsci}      % mathematics and physical sciences
%\bibliographystyle{spphys}       % APS-like style for physics
%\bibliography{}   % name your BibTeX data base

% Non-BibTeX users please use

\end{document}